\documentclass[lettersize,journal]{IEEEtran}

\usepackage{amsmath,amsfonts}
\usepackage{array}
\usepackage[caption=false,font=normalsize,labelfont=sf,textfont=sf]{subfig}
\usepackage{textcomp}
\usepackage{stfloats}
\usepackage{url}
\usepackage{verbatim}
\usepackage{graphicx}
\usepackage{cite}
\usepackage{graphicx}

\usepackage{latexsym}
\usepackage{booktabs}
\usepackage{multirow}
\usepackage{svg}
\usepackage{amsthm,amsmath,amssymb}
\usepackage{mathrsfs}
\usepackage{tabularx}
\usepackage{enumitem}

\begin{document}
\newcolumntype{L}[1]{>{\raggedright\arraybackslash}p{#1}}
\newcolumntype{C}[1]{>{\centering\arraybackslash}p{#1}}
\newcolumntype{R}[1]{>{\raggedleft\arraybackslash}p{#1}}

\title{Intent Propagation Contrastive Collaborative Filtering}

\author{ Haojie Li, Junwei Du$^*$, Guanfeng Liu$^*$, ~\IEEEmembership{Senior Member,~IEEE}, Feng Jiang, Yan Wang, ~\IEEEmembership{Senior Member,~IEEE} and Xiaofang Zhou, ~\IEEEmembership{Fellow,~IEEE}

        % <-this % stops a space
\thanks{Haojie Li, Junwei Du and Feng Jiang are with the School of Data Science, Qingdao University of Science and Technology, Qingdao, China. \protect\\
E-mail: \{lihaojie,dujunwei,jiangfeng\}@qust.edu.cn}% <-this %  }stops a space a space

\thanks{Guanfeng Liu and Yan Wang are with the School of Computing, Macquarie University, Sydney, NSW 2109, Australia.\protect\\ 
E-mail: \{guanfeng.liu, yan.wang\}@mq.edu.au}

% \thanks{Feng Jiang is with the College of Information Science and Technology, Qingdao University of Science and Technology, Qingdao, China. \protect\\
% E-mail: jiangfeng@qust.edu.cn}

% \thanks{Feng Jiang is with the College of Information Science and Technology, Qingdao University of Science and Technology, Qingdao, China. \protect\\
% E-mail: jiangfeng@qust.edu.cn}

\thanks{Xiaofang Zhou is with the Department of Computer Science and Engineering, The Hong Kong University of Science and Technology, Hong Kong SAR, China.
E-mail: zxf@cse.ust.hk}

\thanks{Corresponding authors: Junwei Du and Guanfeng Liu.}


}

\markboth{Journal of \LaTeX\ Class Files,~Vol.~14, No.~8, August~2021}%
{Shell \MakeLowercase{\textit{et al.}}: A Sample Article Using IEEEtran.cls for IEEE Journals}


\maketitle

\begin{abstract}
Disentanglement techniques used in collaborative filtering uncover interaction intents between nodes, improving the interpretability of node representations and enhancing recommendation performance.
However, existing disentanglement methods still face the following two problems.
(1)
They focus on local structural features derived from direct node interactions, overlooking the comprehensive graph structure, which limits disentanglement accuracy.
(2) 
The disentanglement process depends on backpropagation signals derived from recommendation tasks, lacking direct supervision, which may lead to biases and overfitting.
To address the issues, 
we propose the \textbf{I}ntent \textbf{P}ropagation \textbf{C}ontrastive \textbf{C}ollaborative \textbf{F}iltering (IPCCF) algorithm. 
Specifically, we design a double helix message propagation framework to more effectively extract the deep semantic information of nodes, thereby improving the model's understanding of interactions between nodes.
An intent message propagation method is also developed that incorporates graph structure information into the disentanglement process, thereby expanding the consideration scope of disentanglement.
In addition, contrastive learning techniques are employed to align node representations derived from the structure and intents, providing direct supervision for the disentanglement process, mitigating biases, and enhancing the model's robustness to overfitting.
The experiments on three real data graphs illustrate the superiority of the proposed approach.
% The code of IPCCF is available
% at 
% https://github.com/rookitkitlee/IPCCF.
\end{abstract}


\section{Introduction}

% \subsection{Background}

\IEEEPARstart{R}{ecommendation} systems effectively mitigate the challenge of information overload by accurately predicting user-specific preferences \cite{wuSurveyAccuracyorientedNeural2022}. 
Model-based collaborative filtering (CF) methods \cite{yuAreGraphAugmentations2022,maoUltraGCNUltraSimplification2023,caiLightGCLSimpleEffective2023} have become the mainstream recommendation techniques, which generate representations of nodes by mining the interactions between users and items. 
Driven by advancements in deep learning, the development of CF has evolved from shallow models \cite{heFastMatrixFactorization2017,chengAspectAwareLatentFactor2018} to deep models \cite{maoUltraGCNUltraSimplification2023,caiLightGCLSimpleEffective2023}, especially those incorporating Graph Convolutional Networks (GCNs) \cite{heLightGCNSimplifyingPowering2020,wangNeuralGraphCollaborative2019}. 
Models based on GCNs can enhance the understanding of complex user-item interactions, thereby boosting recommendation performance\cite{chenRevisitingGraphBased2020,wuSurveyAccuracyorientedNeural2022}.

However, due to the diversity of user preferences, the interactions between users and items are highly entangled based on various intents \cite{renDisentangledContrastiveCollaborative2023}. 
% If the model does not disentangle these highly entangled interactions, it will lead to suboptimal node representations
% \cite{maDisentangledGraphConvolutional,renDisentangledContrastiveCollaborative2023}. 
A user may interact with various types of items based on differing intents. 
If these intents are ignored in node representations, it will result in high similarity among various types of items, significantly impacting the quality of the representations \cite{maDisentangledGraphConvolutional,renDisentangledContrastiveCollaborative2023}. 
To address this issue, a wide variety of disentangled methods have been proposed \cite{maDisentangledGraphConvolutional,zhaoMultiviewIntentDisentangle2022,liDisentangledContrastiveLearning}. 
These methods concentrate on disentangling the representations of users' latent intents from implicit feedback. 
Their goal is to capture finer-grained interaction patterns between users and items.
To be specific,
DisenGCN \cite{maDisentangledGraphConvolutional} and DGCF \cite{wangDisentangledGraphCollaborative2020} slice user and item representations into chunks, associates each chunk with a latent intent, and iteratively refine the representations of these chunks.
% MIDGN \cite{zhaoMultiviewIntentDisentangle2022}, 
DGCL \cite{liDisentangledContrastiveLearning},
and DCCF \cite{renDisentangledContrastiveCollaborative2023}  utilize contrastive learning methods to disentangle node representations from various perspectives. 
Their objective is to increase disentanglement accuracy and diminish the impact of noise.
KDR \cite{muKnowledgeGuidedDisentangledRepresentation2022} and DIKGNN \cite{tuDisentangledInterestImportance2023} utilize semantic information from knowledge graphs to guide disentangled intent information for interpretable disentanglement.

% \subsection{Motivations}

% The existing disentanglement methods are facing the following two problems that affect their performance in recommendations.
% \textcolor{blue}{
% Existing disentanglement methods still face two issues impacting recommendation performance.}
Although existing disentanglement methods have achieved significant results, there are still two problems that affect their performance in recommendations.

\noindent\textbf{Problem 1.}  
Existing disentanglement methods \cite{wangDisentangledGraphCollaborative2020,renDisentangledContrastiveCollaborative2023,zhang2024exploring} primarily focus on the local structural features formed by direct interactions between
nodes. 
On the one hand, existing methods \cite{wangDisentangledGraphCollaborative2020,zhang2024exploring} rely solely on semantic information extracted from direct interaction relations through GCNs to achieve the disentangling process.
% On the one hand, existing methods \cite{wangDisentangledGraphCollaborative2020,zhang2024exploring} rely solely on semantic information extracted from direct interaction relations to achieve the disentangling process.
% They generate node representations through GCNs.
However, in real-world recommendation scenarios, there exist complex deep semantic relations between nodes. 
These relations not only cover direct interactions but also include high-order interactions and more complex composite interactions between direct and high-order interactions. 
In addition, due to the layer limitations of GCNs during message propagation, the semantic information extracted by GCNs can only reflect local structural information in the graph \cite{xiaHypergraphContrastiveCollaborative2022}.
% Existing disentangle methods \cite{maDisentangledGraphConvolutional, wangDisentangledGraphCollaborative2020,renDisentangledContrastiveCollaborative2023} 
% mainly
% rely on representations generated by GCNs through direct
% interactions between nodes. 
% Due to the layer limitations of GCNs during message propagation, these representations primarily reflect local structural information in the graph, making it difficult to capture more comprehensive structural details\cite{xiaHypergraphContrastiveCollaborative2022}.
% Therefore, the information relied upon by existing disentanglement algorithms \cite{maDisentangledGraphConvolutional, wangDisentangledGraphCollaborative2020,renDisentangledContrastiveCollaborative2023} only reflect local structural information in the graph, making it difficult to capture more comprehensive structural details.
The above two drawbacks both limit the model's understanding of node interaction intents, potentially leading to misunderstandings or biases during the disentanglement process.
On the other hand, existing methods \cite{wangDisentangledGraphCollaborative2020,renDisentangledContrastiveCollaborative2023}  only focus on disentangling the interaction intents of direct interactions between users and items. 
They do not examine the rationality of the disentangled results from a comprehensive graph structure perspective.
This ultimately leads to the overfitting of models, since these models strive to achieve overall optimization goals.
% Not only does this impact the quality of disentanglement, but it could also negatively influence the outcomes of subsequent recommendation tasks.

\noindent\textbf{Problem 2.} 
% Due to the lack of direct supervision signals during the disentanglement process, existing disentanglement methods \cite{maDisentangledGraphConvolutional, wangDisentangledGraphCollaborative2020} are prone to biases and overfitting.
Existing disentanglement methods \cite{maDisentangledGraphConvolutional, wangDisentangledGraphCollaborative2020,zhang2024exploring} lack direct supervision signals during the disentanglement process.
% This leads to bias and overfitting in the
% model.
They improve the effectiveness of node representations by carefully exploring the interaction intents between nodes.
However, the intents behind user-item interactions are latent variables that cannot be directly observed from the data.
Existing models \cite{maDisentangledGraphConvolutional, zhaoMultiviewIntentDisentangle2022} rely on indirect supervision signals obtained from the backpropagation process of the recommendation task to guide the disentanglement process.
Under these supervision signals, the goal of disentanglement is to ensure that the final node representations meet the requirements of the recommendation task, rather than accurately reflecting the actual interaction intents between nodes. 
Therefore, excessive optimization for recommendation objectives may lead to biases in the disentanglement process and the overfitting of models.
In addition, although iterative algorithms like Expectation Maximization \cite{congBigLearningExpectation2024} can estimate the values of latent variables, their strong dependence on initial settings may cause the model to converge to suboptimal solutions.
% In addition, iterative algorithms, due to the need to progressively solve latent variables and optimize the recommended objectives, may lead to slow convergence speeds and high training costs.

To address the above problems, 
we propose the \textbf{I}ntent \textbf{P}ropagation \textbf{C}ontrastive \textbf{C}ollaborative \textbf{F}iltering (IPCCF) algorithm. 
Specifically, we extract high-order interaction relations between homogeneous nodes to enhance the capture of graph structural features.
A double helix message propagation framework is designed to better integrate direct and high-order interactions between nodes.
By alternately executing different types of message propagation processes, IPCCF generates node representations containing deep semantic information, enhancing the model's understanding of interaction intents between nodes.
Additionally, we design an intent message propagation method based on the interaction intents between nodes. 
It aggregates node intent representations via graph structures, integrating the graph's structural information into the disentanglement process. 
This broadens the consideration scope of disentanglement and examines the rationality of the results from the perspective of comprehensive graph structure.
Furthermore, we use contrastive learning techniques to align the node representations generated by message propagation based on structures with those generated by message propagation based on intents.
This provides direct supervision signals for the disentanglement process, reducing biases and enhancing the model's resistance to overfitting.
To sum up, the contributions of our work are summarized as follows:

\begin{itemize}[leftmargin=10pt]

\item
% We propose a new double helix message propagation framework to extract deep semantic information of nodes, which can enhance the model's understanding of interaction intents between nodes.
% An intent message propagation mechanism is proposed to enhance the model's disentanglement process by evaluating the rationality of results from a more comprehensive graph structure perspective.
% An intent message propagation mechanism is introduced to broaden the consideration scope of the model's disentanglement process. 
% It examines the rationality of the disentangled results from a more comprehensive graph structure perspective.
We propose a double helix message propagation framework to extract deep semantic node information, improving the model's understanding of interaction intents. Additionally, an intent message propagation mechanism is proposed to enhance the disentanglement process by evaluating results from a comprehensive graph structure perspective.

\item 
% We employ a contrastive learning approach to align node representations generated by different message propagation strategies, leveraging the structural relations between nodes to provide direct supervision signals for the disentanglement process. 
% This reduces biases in the disentanglement process and enhances the model's resistance to overfitting.
We use contrastive learning to align node representations from different propagation strategies, leveraging structural relations to supervise disentanglement, reducing biases, and improving resistance to overfitting.

\item We conduct experiments on three real-world datasets. 
Compared to the best-performing baseline, our model achieves an average improvement of 
$6.37\%$ in Precision,
$6.14\%$ in Recall and $9.36\%$ in NDCG, respectively.

% Compared to the best-performing baseline, our model achieves an average improvement of 
% \textcolor{blue}{$6.37\%$ in the Precision metric,}
% $6.14\%$ in the Recall metric and $9.36\%$ in the NDCG metric, respectively.
% Compared to the best results achieved by other methods, our model demonstrates an average improvement of $6.14\%$ in the Recall metric and $9.36\%$ in the NDCG metric.

\end{itemize}

\section{Related work}

\subsection{GCN-based Recommendation Models}

% The application of Graph Neural Networks (GNNs) in CF tasks has demonstrated their remarkable ability to learn representations of graph-structured user-item interaction data \cite{zhuAdaMCLAdaptiveFusion2023}. 
% It is worth noting that GCNs, as a widely used variant of GNNs, have further promoted the development of graph-based recommendation systems \cite{kipfSEMISUPERVISEDCLASSIFICATIONGRAPH2017}. 
% GCN-based methods can uncover complex interaction relations between nodes by recursively performing message propagation on graph structures \cite{chenHeterogeneousGraphContrastive2023,yangEnhancedGraphLearning2021}.
% Therefore, some studies \cite{wangNeuralGraphCollaborative2019,zhangSTARGCNStackedReconstructed2019} are dedicated to developing efficient message propagation mechanisms to enhance recommendation performance.

GCN-based methods have been proposed to provide effective message propagation, enhancing recommendation performance.
For example, NGCF \cite{wangNeuralGraphCollaborative2019} focuses on the high-order connectivity between nodes during the message propagation process; 
STAR-GCN \cite{zhangSTARGCNStackedReconstructed2019} significantly improves prediction performance by combining a series of GCN encoders and decoders with intermediate supervision strategies. 
% GC-MC \cite{bergGraphConvolutionalMatrix2017} enriches the message propagation process in graph neural networks from the perspective of link prediction.
In addition, some studies \cite{chenRevisitingGraphBased2020,heLightGCNSimplifyingPowering2020} aim to simplify the GCN architecture to improve the efficiency of model training. 
For example, LR-GCCF \cite{chenRevisitingGraphBased2020} achieves both performance improvement and complexity reduction in collaborative filtering tasks by eliminating non-linear transformations; 
LightGCN \cite{heLightGCNSimplifyingPowering2020} has shown that feature transformation and non-linear activation in GCNs may be unnecessary for collaborative filtering, paving the way for simplified model structures.
Although GCN-based recommendation models are effective, they fail to address data sparsity and the entangled interaction intent problems in real-world recommendation scenarios.
This makes it difficult to accurately capture user preferences.
% Although GCN-based recommendation models are effective, the limited number of message propagation layers poses a constraint \cite{xiaHypergraphContrastiveCollaborative2022}. 
% This constraint causes nodes to only contain local graph information, which hinders the performance of subsequent tasks.

% Although GCN-based recommendation models exhibit significant effectiveness, the sparsity of data still imposes limitations on their performance
% \cite{zhuAdaMCLAdaptiveFusion2023}. 

\subsection{Contrastive Learning Recommendation Models}
 
% Contrastive learning learns representational invariances through data augmentation, thereby reducing the impact of data sparsity and noise on model performance \cite{wuSelfsupervisedGraphLearning2021,guoLGMRecLocalGlobal2024,liuGraphDisentangledContrastive2024}.
Generally, contrastive learning first generates contrasting views through data augmentation, and then maximizes mutual information to encourage consistency between different contrasting views.
For example, SGL \cite{wuSelfsupervisedGraphLearning2021} generates multiple views of a node through data augmentation on the graph structure, aiming to maximize consistency between different views of the same node and minimize consistency between views of different nodes. 
SimGCL \cite{yuAreGraphAugmentations2022} abandons graph augmentation and instead manipulates the uniformity of learned representations by introducing consistent noise into the embedding space, thereby generating contrasting views.
DGI \cite{velickovicDeepGraphInfomax2018}, RGCL \cite{shuaiReviewawareGraphContrastive2022}, and LGMRec \cite{guoLGMRecLocalGlobal2024} enhance the effectiveness of node representations by constructing augmented views with both local and global features.
BiGI \cite{cao2021bipartite} 
aggregates information from 2-hop neighbors to address the different node types in bipartite graphs and enhance global relevance by aligning local and global representations.
NCL \cite{linImprovingGraphCollaborative2022} explicitly incorporates potential neighbors into contrast pairs from both structural and semantic perspectives. 
% HCCF \cite{xiaHypergraphContrastiveCollaborative2022} and AdaCML \cite{zhuAdaMCLAdaptiveFusion2023} incorporate high-order message propagation relations between nodes into the contrastive learning framework.
% HCCF \cite{xiaHypergraphContrastiveCollaborative2022} employs a hypergraph-enhanced cross-view contrastive learning framework to jointly capture local and global collaborative relations. 
SimRec \cite{xiaGraphlessCollaborativeFiltering2023} combines knowledge distillation and contrastive learning to enable adaptive knowledge transfer from the teacher model to lightweight student networks. 
Although contrastive learning recommendation models enhance node representation through augmented views, these methods overlook the interaction intents between nodes. 
Ignoring these intents when generating node representations could lead to suboptimal solutions for the model \cite{maDisentangledGraphConvolutional,renDisentangledContrastiveCollaborative2023}.

\subsection{Disentangled Recommendation Models}

% In recommendation systems, users may interact with different categories of items due to various intents \cite{renDisentangledContrastiveCollaborative2023,wangDisentangledGraphCollaborative2020}.
Learning disentangled representations of users' latent intents can refine user preferences, and enhance the accuracy of recommendation systems \cite{maDisentangledGraphConvolutional,zhaoMultiviewIntentDisentangle2022,wangDisentangledGraphCollaborative2020}.
For example,
DisenGCN \cite{maDisentangledGraphConvolutional} and DGCF \cite{wangDisentangledGraphCollaborative2020} 
divide the interaction relations between nodes according to interaction intents, obtaining the representation of nodes under each intent;
MIDGN \cite{zhaoMultiviewIntentDisentangle2022} 
and LGD-GCN \cite{guoLGDGCNLocalGlobal2023}
focus on disentangling user interaction intents from both global and local perspectives;
KDR \cite{muKnowledgeGuidedDisentangledRepresentation2022} and DIKGNN \cite{tuDisentangledInterestImportance2023}  use semantic information from knowledge graphs to guide the disentanglement of interaction intents between nodes, achieving interpretable disentanglement;
GDCF \cite{zhangGeometricDisentangledCollaborative2022} infers concepts linked to user intents and various geometric shapes to generate disentangled geometric representations;
DCCF \cite{renDisentangledContrastiveCollaborative2023} achieves adaptive intent disentanglement through self-supervised augmentation, and extracts finer-grained latent factors via global context learning;
GDCCDR \cite{liuGraphDisentangledContrastive2024} proposes an adaptive, parameter-free filter to measure the importance of various interactions. This helps create more detailed disentangled representations.
Although existing disentangled recommendation models have achieved good results, they primarily focus on local structural features formed by direct interactions between nodes. 
They fail to consider the disentanglement process from a more comprehensive graph structure perspective. Moreover, these disentanglement methods lack effective supervision signals during the process. 
This may lead to biases in the disentanglement process and the overfitting of models.

\subsection{High-Order Relation Extraction Methods}
% \textcolor{blue}{
% When generating node representations in recommendation models, considering high-order relations between nodes can enhance the model's understanding of interaction relations. 
% This improves the quality of node representations and ultimately boosts recommendation performance.
% There have been many studies \cite{gao2020learning,li2021hyperbolic}  on extracting higher-order relations between nodes.
% Random walk is a commonly used technique for capturing high-order relations between nodes and has been widely applied in various homogeneous network embedding models \cite{perozzi2014deepwalk,grover2016node2vec}.
Considering high-order relations between nodes in recommendation models can enhance interaction understanding and improve node representation quality.
There have been many studies on high-order relation extraction \cite{gao2020learning,sybrandt2019first,li2021hyperbolic}.
For example,
BiNE \cite{gao2020learning} employs a biased generator based on the random walk method \cite{perozzi2014deepwalk} to preserve long-tail distribution characteristics in generated node sequences. It then extracts high-order relations between nodes from the sequences.
FOBE \cite{sybrandt2019first} defines relations between homogeneous nodes by sharing at least one common neighbor; HOBE \cite{sybrandt2019first}  refines this by using algebraic distance to measure the strength of node connectivity.
HNCR \cite{li2021hyperbolic}, based on the extraction of homogeneous node structural relations, further derives the node embeddings to capture the semantic relations between nodes.
DICER \cite{fu2021dual} measures the proportion of shared neighbors between two nodes and selects node pairs that exceed a specific threshold as the high-order relations to be extracted.
Although the above high-order relation extraction methods \cite{gao2020learning,fu2021dual,li2021hyperbolic} perform well in their tasks, their primary goal is to enhance node representation. 
% Therefore, their selection should align with the needs of subsequent models. As these needs differ from our model's, directly copying them won’t yield optimal results.
Therefore, the model should select the high-order relation extraction methods based on subsequent requirements, rather than solely relying on their performance in the original task.

\begin{figure*}[h]
\setlength{\abovecaptionskip}{0pt}
\centering
% \includesvg[scale=0.34]{Figure/F2.svg}
\includegraphics[scale=0.34]{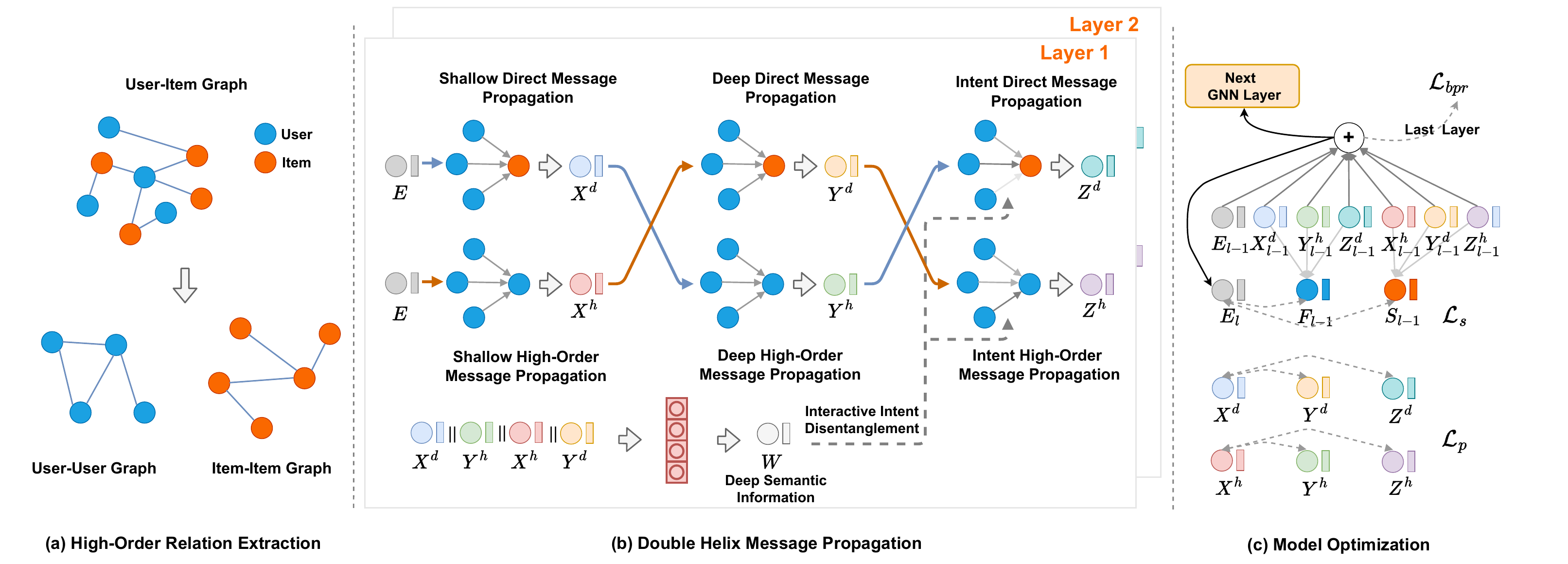}
\caption{
% An overview of IPCCF.
The overall framework of the IPCCF model includes three key modules: the high-order relation extraction module, which better captures the structural features of the graph; the dual-helix message propagation module, which generates node representations by integrating structure and intent; and the model optimization module, which aligns node representations from multiple perspectives to enhance their quality.
}
\label{fig:label}
\end{figure*}

\begin{table}
\setlength{\abovecaptionskip}{5pt}
\setlength{\belowcaptionskip}{-5pt}
  \centering
  \scriptsize
  \caption{NOTATIONS AND EXPLANATIONS.}
  % \resizebox{\linewidth}{!}{
  \resizebox{8cm}{!}{
    \begin{tabular}{L{1.2cm}L{6.2cm}}
    \toprule

    Notation   & Explanation       \\  \midrule

    $\mathcal{U}$ and $\mathcal{I}$  &
    the user and item sets   \\ 
    $\mathcal{A}$  & the observed implicit feedback matrix   \\
    $\mathcal{N}(u)$ &  the set of direct neighbors of node $u$   \\
    $\eta$ and $Q$ & the filter thresholds for extracting high-order relations between nodes \\
    
    % $\mathcal{H}^{(u)}$ and $\mathcal{H}^{(i)}$  & 
    % the similarity matrices for user-user and item-item \\

    $\mathcal{H}$ & 
    the similarity matrix between homogeneous nodes \\

    $\hat{\mathcal{A}}$ and 
    $\hat{\mathcal{H}}$ &
    the direct and high-order interaction matrices
    \\
    % $\mathcal{N}_{\hat{\mathcal{A}}}(i)$ and $\mathcal{N}_{\hat{\mathcal{H}}}(i)$ &
    % the sets of first-order neighbors of node $i$ in $\hat{\mathcal{A}}$ and $\hat{\mathcal{H}}$
    % \\
    $\textbf{F}$ and $\textbf{S}$ &
    the first and second helical sequences
    \\
    $\textbf{E}$ &
    the embedding matrices of nodes \\
    $d$ &  the dimension of the embeddings \\

    $\bar{\mathcal{A}}$ and 
    $\bar{\mathcal{H}}$ &
    the normalized direct and high-order interaction matrices
    \\

    $\textbf{X}^{d}$ and $\textbf{X}^{h}$ &
    the node representations generated through shallow direct and high-order message propagation \\

    $\textbf{Y}^{d}$ and $\textbf{Y}^{h}$ &
    the node representations generated through deep direct and high-order message propagation \\

    $\textbf{Z}^{d}$ and $\textbf{Z}^{h}$ &
    the node representations generated through intent direct and high-order message propagation \\

    ${\mathcal{T}}$ &
    the mask matrix that reflects the interaction intensity between nodes \\

    $\tilde{\mathcal{R}}$ & 
    the learnable intensity matrix derived from ${\mathcal{T}}$ \\

    $\ddot{\mathcal{R}}^{k}_{u,v}$ & the weight of intent $k$ in the propagation process between node $u$ and node $v$ \\

    $\Bar{\mathcal{R}}$ & the normalized intent message propagation matrix \\

    $\textbf{C}$ &
    the embedding matrices of intents \\

    $K$ & the number of intents \\ 

    $\textbf{W}$ &
    the node representations that can reflect deep semantic information \\

    $\textbf{M}$ &
    the node representation before intent message propagation \\

    $\textbf{N}$ &
    the node representation after intent message propagation \\

    $L$ & the layers of message propagation \\

    $\textbf{E'}$ & 
    the final embedding matrices for nodes \\

    $\hat{\mathcal{R}}_{u,i}$ & the prediction score between user $u$ and item $i$ \\

    $\Vert \cdot \Vert_{2}$ & the L2-norm method \\

    $\Vert \cdot \Vert^2_F$
    & the Frobenius norm method \\

   \bottomrule 

    \end{tabular}

   }
\end{table}
\section{METHODOLOGY}

We define the user set as $\mathcal{U}=\{ u \}$
and the item set as 
$\mathcal{I} = \{ i \}$. 
The observed implicit feedback matrix is denoted by
$\mathcal{A} \in \mathbb{R}^{|\mathcal{U}| \times |\mathcal{I}|}$, 
where 
$\mathcal{A}_{u,i} = 1 $ if user 
$u$ has interacted with item $i$, and 
$\mathcal{A}_{u,i} = 0 $ otherwise.
The notations used in this paper are shown in Table I.

The overall framework of IPCCF, as shown in Fig. 1, consists of three modules: 
\begin{itemize}[leftmargin=10pt]
\item \textbf{The Module of High-Order Relation Extraction}:
% \textcolor{blue}{
This module extracts high-order relations among homogeneous nodes to enrich GCN propagation, improving the capture of graph structural features.
% In this module, we extract high-order relations among homogeneous nodes to enrich the GCN message propagation process. 
% This helps us better capture the structural features of the graph.
\item \textbf{The Module of Double Helix Message Propagation}:
This module introduces a double helix message propagation framework to enrich node semantic representation and improve interaction understanding. An intent message propagation mechanism is proposed to enhance disentanglement by evaluating results from a graph structure perspective.
% In this module, 
% we propose a double helix message propagation framework to enrich the semantic representation of nodes, enhancing the model's understanding of interactions between nodes. An intent message propagation process is also introduced, expanding the consideration scope of disentanglement. 
% In addition, disentangled node representations are also integrated into this message propagation framework, further enhancing the semantic accuracy of the recommendation system.
\item \textbf{The Module of Model Optimization}:
This module integrates node representations from different message propagations to predict user preferences accurately. It also uses contrastive learning to align these representations, providing supervision for disentanglement, thereby improving accuracy and enhancing resistance to overfitting.
% In this module, we integrate node representations generated from different message propagations to accurately predict user preferences. 
% In addition, we use contrastive learning methods to align these node representations, providing direct supervision signals for the disentanglement process. 
% This improves the accuracy of disentanglement and enhances the
% model’s resistance to overfitting.

\end{itemize}

% In this section, we will introduce each module in detail.
% Before introducing each module in detail, we will first introduce the intent propagation learning paradigm we proposed to address the efficiency issues of the disentangling method.

\subsection{High-Order Relation Extraction}

Due to the limitations of the message propagation layers, node representations based on direct interaction relations can only reflect the local structural features of the graph \cite{xiaHypergraphContrastiveCollaborative2022}. 
Furthermore, in models that use graph convolution to generate node representations, nodes often suffer from over-smoothing as a result of absorbing too many signals from their direct neighbors \cite{linImprovingGraphCollaborative2022, xiaHypergraphContrastiveCollaborative2022, zhuAdaMCLAdaptiveFusion2023}. 

To address this issue, we explore high-order interactions between nodes.
In our model, the 2-hop relations between nodes are referred to as high-order relations. This definition is consistent with the one used in AdaCML \cite{zhuAdaMCLAdaptiveFusion2023}. Similarly, in CSE \cite{chen2019collaborative} and NCL \cite{linImprovingGraphCollaborative2022}, relations beyond 1-hop are also referred to as high-order relations.
% Through high-order message propagation, node representations can acquire more contrastive signals, thereby effectively enhancing the quality of node representations. 
Since high-order relation extraction is not the focus of this work, we adopt the appropriate methods from existing studies \cite{gao2020learning,li2021hyperbolic,fu2021dual} based on the following requirements.
(1) We need to extend the original direct relations with high-order relations to enhance the model's ability to perceive the graph structure. 
(2) Our model will disentangle the interactions between nodes based on intent. We should minimize the number of high-order relations to maintain efficiency.
Based on the above requirements, HNCR \cite{li2021hyperbolic} extracts semantic neighbors, which cannot supplement structural relations.
The high-order relation extraction methods in BiNE \cite{gao2020learning}, FOBE \cite{sybrandt2019first}, and HOBE \cite{sybrandt2019first} lack pruning capabilities, while the relation pruning process of DICER \cite{fu2021dual} is too coarse to ensure the completeness of the extracted relations. 
% In order to meet the above requirements, we use the Jaccard Similarity Coefficient method to extract high-order relations between nodes from the structure. 
% By adjusting two hyperparameters, $\eta$ and $Q$, this method ensures the minimal complete subset of high-order relations and has also been applied in the AdaCML \cite{zhuAdaMCLAdaptiveFusion2023}.
Given these shortcomings, we use the Jaccard Similarity Coefficient method to calculate the similarity between homogeneous nodes. This method effectively measures the proportion of shared neighbors between two nodes, enhancing the model’s ability to perceive the graph structure. 
By adjusting two hyperparameters, $\eta$ and $Q$, we can minimize the number of high-order relations between nodes while ensuring model performance, thus improving training efficiency.
This method is also applied in AdaCML \cite{zhuAdaMCLAdaptiveFusion2023}.
We define the collaborative similarity relation between the user (item) $u$ and the user (item) 
$v$ as
% The collaborative similarity relations between high-order nodes can be calculated based on their shared interaction history \cite{wuDualGraphAttention2019,fuDualSideDeep2021}. 
% In IPCCF, we define the collaborative similarity relation between the user (item) $u$ and the user (item) 
% $v$ through the Jaccard Similarity Coefficient.
\begin{equation}
\small
sim_{u,v} = \frac{|\mathcal{N}(u) \cap \mathcal{N}(v) |}{|\mathcal{N}(u) \cup \mathcal{N}(v)|},
\end{equation}
where $\mathcal{N}(u)$ represents the set of direct neighbors of node 
$u$.
% To improve computational efficiency by filtering out weaker propagation signals among homogenous nodes, we define the similarity matrix between homogenous nodes as follows:
To improve computational efficiency, we filter out weaker propagation signals between homogeneous nodes. 
The similarity matrix between homogeneous nodes is defined as
\begin{equation}
\small
\mathcal{H}_{u,v} = 
\begin{cases}
sim_{u,v}, & \text {$sim_{u,v} \ge \eta$ or top $Q$ values for node $u$;} \\
0, & \text {otherwise.}
\end{cases}
\end{equation}
We use 
$\mathcal{H}^{(u)} \in \mathbb{R}^{|\mathcal{U}| \times |\mathcal{U}|}$ and $\mathcal{H}^{(i)} \in \mathbb{R}^{|\mathcal{I}| \times |\mathcal{I}|}$ to represent the similarity matrices for user-user and item-item, respectively.

\subsection{Double Helix Message Propagation}

Double helix message propagation framework consists of three components: shallow message propagation, deep message propagation, and intent message propagation.
Among them, shallow message propagation and deep message propagation are established based on the interaction structures between nodes, while intent message propagation is established based on the interaction intents between nodes.
The message propagation process in each component is designed from direct and high-order relations between nodes.
Let 
$\hat{\mathcal{A}}$ and 
$\hat{\mathcal{H}}$  denote the direct and high-order interaction matrices, respectively.
\begin{equation}
\small
\hat{\mathcal{A}} = 
\begin{pmatrix}
0 & \mathcal{A} \\
{\mathcal{A}}^T & 0
\end{pmatrix} \;
\hat{\mathcal{H}} = 
\begin{pmatrix}
\mathcal{H}^{(u)} & 0 \\
0 & \mathcal{H}^{(i)}
\end{pmatrix}
\end{equation}
Let $\mathcal{N}_{\hat{\mathcal{A}}}(i)$ and $\mathcal{N}_{\hat{\mathcal{H}}}(i)$ be the sets of first-order neighbors of node $i$ in $\hat{\mathcal{A}}$ and $\hat{\mathcal{H}}$, respectively.
Information for two helix sequences is stored in $\textbf{F} \in \mathbb{R}^{(|\mathcal{U}| + |\mathcal{I}|) \times d}$ 
and $\textbf{S} \in \mathbb{R}^{(|\mathcal{U}| + |\mathcal{I}|) \times d}$, respectively.
% We use 
% $\textbf{F} \in \mathbb{R}^{(|\mathcal{U}| + |\mathcal{I}|) \times d}$ 
% and $\textbf{S} \in \mathbb{R}^{(|\mathcal{U}| + |\mathcal{I}|) \times d}$ to store information for two helix sequences, respectively.

\subsubsection{Shallow Message Propagation}

In the shallow message propagation process, we construct message propagation methods based on both direct and high-order interactions among nodes to achieve shallow aggregation of neighborhood information.

\noindent\textbf{Shallow Direct Message Propagation.}
GNNs have become a popular learning method for capturing CF signals in recommendation systems \cite{zhuAdaMCLAdaptiveFusion2023}. 
In IPCCF, we adopt a method similar to LightGCN \cite{heLightGCNSimplifyingPowering2020} to implement the shallow direct message propagation.
% The specific process is as follows:
% \begin{equation}
% \small
% \textbf{X}^{d} = \bar{\mathcal{A}} \cdot \textbf{E}, 
% \end{equation}
The process is
$\textbf{X}^{d} = \bar{\mathcal{A}} \cdot \textbf{E}$,
where $\textbf{E} \in \mathbb{R}^{(|\mathcal{U}| + |\mathcal{I}|) \times d}$ represents the embedding matrices for nodes, and $d$ denotes the dimension of the embeddings. 
The aggregated representations obtained through shallow direct message propagation are denoted as $\textbf{X}^{d} \in \mathbb{R}^{(|\mathcal{U}| + |\mathcal{I}|) \times d}$. 
% $\bar{\mathcal{A}} \in \mathbb{R}^{(|\mathcal{U}| + |\mathcal{I}|) \times (|\mathcal{U}| + |\mathcal{I}|) }$ 
$\bar{\mathcal{A}}$ 
serves as the normalized adjacency matrix derived from the direct interaction matrix 
$\hat{\mathcal{A}}$, which is defined as
\begin{equation}
\small
\bar{\mathcal{A}}_{u,i}
= \frac{1}
{
\sqrt{|\mathcal{N}_{\hat{\mathcal{A}}}(u)| \times |\mathcal{N}_{\hat{\mathcal{A}}}(i)|}
}.
\end{equation}
After completing the shallow direct message propagation, we update the 1st helical sequence by
\begin{equation}
\small
\textbf{F} = \textbf{X}^{d}.
\end{equation}

% To capture deep collaborative filtering signals, we conduct message propagation based on GNNs across distinct layers.
% The generation process of the 
% $l$-th layer's embedding representation is as follows:
% \begin{equation}
% \small
% \textbf{E}^{(u)}_{l} = \textbf{E}^{(u)}_{l-1} + \textbf{Z}^{(u)}_{l-1}, \;
% \textbf{E}^{(i)}_{l} = \textbf{E}^{(i)}_{l-1} + \textbf{Z}^{(i)}_{l-1}.
% \end{equation}
% To reduce over-smoothing effects, we apply residual connections during the aggregation phase \cite{xiaHypergraphContrastiveCollaborative2022,renDisentangledContrastiveCollaborative2023}.

% \subsubsection{Macro High-Order Message Propagation}

\noindent\textbf{Shallow High-Order Message Propagation.}
The high-order message propagation process among homogeneous nodes is constructed using a method similar to direct message propagation. 
% The specific process is as follows:
% \begin{equation}
% \small
% \textbf{X}^{h} = \bar{\mathcal{H}} \cdot \textbf{E}, 
% \end{equation}
The process is
$\textbf{X}^{h} = \bar{\mathcal{H}} \cdot \textbf{E}$, 
where $\textbf{X}^{h} \in \mathbb{R}^{(|\mathcal{U}| + |\mathcal{I}|) \times d}$ denotes the aggregated representations through their high-order neighbor nodes.
$\bar{\mathcal{H}}$ denotes the normalized matrix derived from the high-order matrix 
$\hat{\mathcal{H}}$, which is defined as
\begin{equation}
\small
\bar{\mathcal{H}}_{u,v} = \frac{
\hat{\mathcal{H}}_{u,v}
}{
\sum_{k \in \mathcal{N}_{\hat{\mathcal{H}}}(u)} \hat{\mathcal{H}}_{u,k}
}.
\end{equation}
After completing the shallow high-order message propagation, we update the 2nd helical sequence by 
\begin{equation}
\small
\textbf{S} = \textbf{X}^{h}.
\end{equation}

\subsubsection{Deep Message Propagation}

To enrich the semantic information of node representations and enhance disentanglement, we optimize the message propagation process in two key ways:
(1) Expanding the scope of node representations to encompass a broader range of structural information.
(2) Integrating various types of message propagation methods more effectively, enabling nodes to better assimilate information from neighbors across different hops.
To implement this, we design a deep message propagation component that employs a sequential approach to combine various message propagation methods, allowing for a deep exploration of the nodes' semantic information.

\noindent\textbf{Deep Direct Message Propagation.}
We implement deep direct message propagation based on the
normalized user-item interaction matrix $\bar{\mathcal{A}}$. 
This propagation process takes the output of shallow high-order message propagation $\textbf{X}^h$ as its input, aiming to expand the scope of node representations and more effectively integrate different message propagation methods.
% The specific process is as follows:
% \begin{equation}
% \small
% \textbf{Y}^d = \bar{\mathcal{A}} \cdot \textbf{X}^{h}, 
% \end{equation}
The process is
$\textbf{Y}^d = \bar{\mathcal{A}} \cdot \textbf{X}^{h}$,
where
$\textbf{Y}^{d} \in \mathbb{R}^{(|\mathcal{U}|+|\mathcal{I}|) \times d}$ represents the node representations generated by deep direct message propagation.
% Then, we update the 2nd helical sequence by the following equation: 
Then, we update the 2nd helical sequence by
\begin{equation}
\small
\textbf{S} = \textbf{X}^{h} + \textbf{Y}^{d}.
\end{equation}

\noindent\textbf{Deep High-Order Message Propagation.}
Similar to the method of deep direct message propagation, 
we use the normalized high-order neighbor matrix 
$\bar{\mathcal{H}}$ to implement the message propagation process, with the input being the node representation 
$\textbf{X}^d$ generated by shallow direct message propagation.
% The specific process is as follows:
% \begin{equation}
% \small
% \textbf{Y}^{h} = \bar{\mathcal{H}} \cdot \textbf{X}^{d}, 
% \end{equation}
The process is
$\textbf{Y}^{h} = \bar{\mathcal{H}} \cdot \textbf{X}^{d}$,
where
$\textbf{Y}^{h} \in \mathbb{R}^{(|\mathcal{U}| + |\mathcal{I}|) \times d}$ represents the node representations generated by deep high-order message propagation.
% Then, we update the 1st helical sequence by the following equation: 
Then, we update the 1st helical sequence by
\begin{equation}
\small
\textbf{F} = \textbf{X}^{d} + \textbf{Y}^{h}.
\end{equation}

\subsubsection{Intent Message Propagation}
To avoid limiting the disentanglement process to direct interactions between users and items, we create intent direct message propagation and intent high-order message propagation based on the interaction intents between nodes. 
% Intent aggregation based on interaction relations is achieved.
Through the intent message propagation, we achieve aggregation of node intent representations via the graph structure. 
This expands the consideration scope of disentanglement and examines the rationality of disentanglement results from a more comprehensive graph structure perspective.
We believe that interactions between nodes are driven by different intents.
Let $\textbf{C} \in \mathbb{R}^{K \times d}$ denote the intent representations hidden in the interactions, 
where $K$ denotes the number of intents.
The probability of user $u$ interacting with item 
$i$ based on intents is defined as 
$\hat{R}_{u,i}=\sum_k^K {c_u^k}^T {c_i^k}$, 
where
${c_x^k}$ represents the representation of node 
$x$ under the 
$k$-th intent.
% We obtain the representation of nodes under each intent through intent message propagation. 
We obtain the intent-based representation of nodes through intent message propagation.
The overall process is as follows and is shown in Fig. 2.
% The overview of intent message propagation is as follows and is shown in Fig. 2.

\begin{figure}[t]
\setlength{\abovecaptionskip}{2pt}
\setlength{\belowcaptionskip}{-5pt}
\centering
% \includesvg[scale=0.28]{Figure/F3.svg}
\includegraphics[scale=0.28]{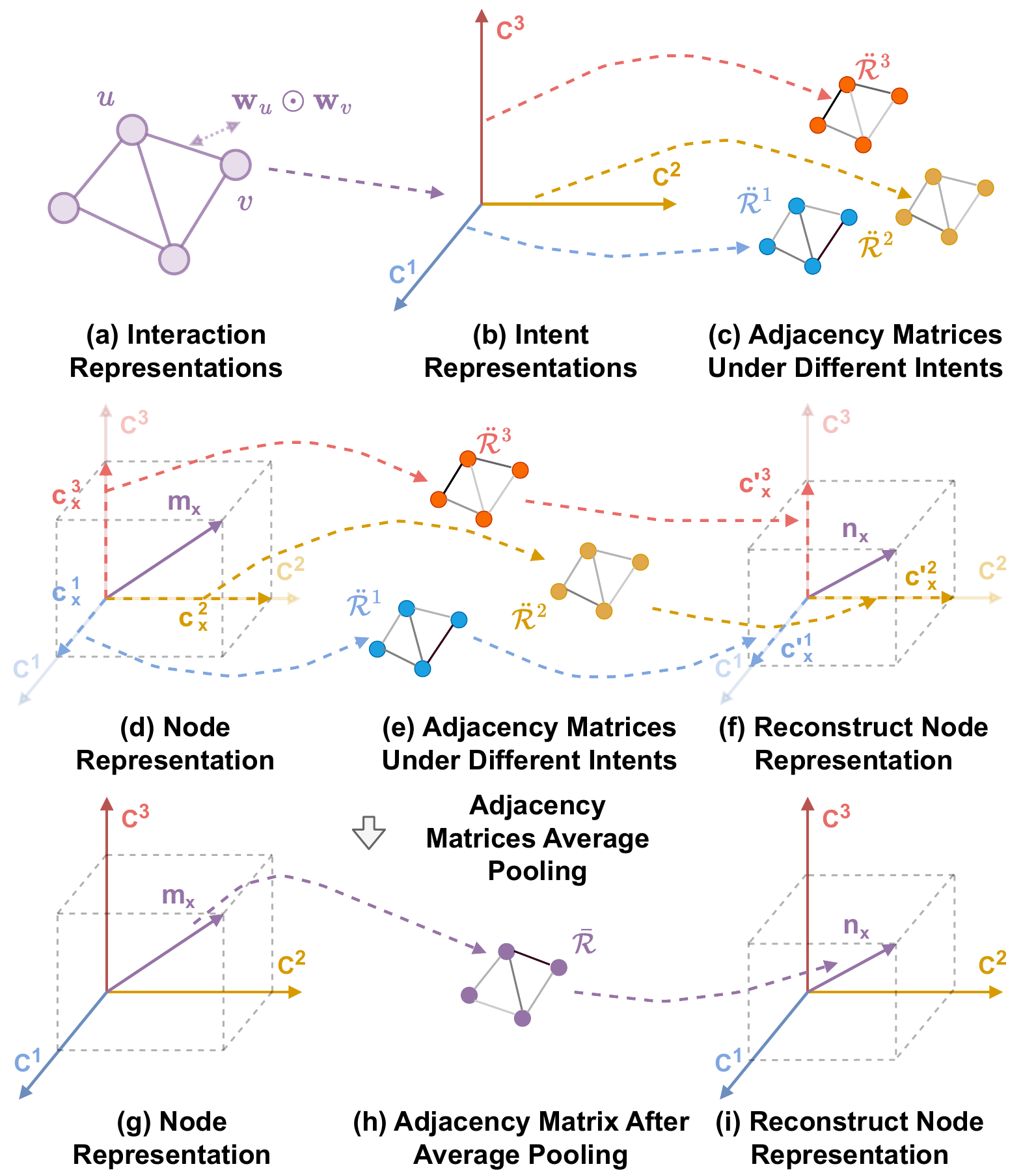}
\caption{Intent message propagation.}
\label{fig:label}
\end{figure}

\begin{itemize}[leftmargin=10pt]
\item We construct interaction representations from node representations, disentangle them by intents, and obtain separate adjacency matrices for each intent, as shown in Fig. 2(a-c).
\item We view the disentangling of node representations as projecting them under different intents, as shown in Fig. 2(d). 
\item 
We perform message propagation on node representations under each intent based on the corresponding adjacency matrices, as shown in Fig. 2(e).
% We construct message propagation matrices for each intent based on the projection strength of nodes under different intents and perform message propagation, as shown in Fig. 2(b). 
\item After intent message propagation, node representations are reconstructed, as shown in Fig. 2(f).
\end{itemize}

In this process, performing message propagation for each intent is inefficient. 
To maximize the disentangling effect, existing models \cite{wangDisentangledGraphCollaborative2020} minimize the similarity between intent representations, making them approximately orthogonal in the latent space, so that their message propagation processes do not affect each other.
In addition, we use LightGCN to implement intent message propagation, simplifying it to matrix-vector multiplication. 
Therefore, by applying average pooling to the adjacency matrices for each intent, we streamline the process and enhance its efficiency, as shown in Fig. 2(g-i).
% Through the intent message propagation, we achieve aggregation of node intent representations via the graph structure. 
% This expands the scope of disentanglement considerations, shifting beyond direct interaction between nodes to incorporate the graph structure into the disentanglement process.

Based on the above discussion, we define the intent message propagation procedure 
$P$. 
The inputs to this procedure include the deep semantic representation $\textbf{W} \in \mathbb{R}^{(|\mathcal{U}| + |\mathcal{I}|) \times d}$ of nodes, the interaction relation 
$\mathcal{R} \in \mathbb{R}^{(|\mathcal{U}| + |\mathcal{I}|) \times (|\mathcal{U}| + |\mathcal{I}|) }$ between nodes, and the node representation $\textbf{M} \in \mathbb{R}^{(|\mathcal{U}| + |\mathcal{I}|) \times d}$ before intent message propagation. 
The output of this procedure is the node representation  $\textbf{N} \in \mathbb{R}^{(|\mathcal{U}| + |\mathcal{I}|) \times d}$ after intent message propagation.
In procedure $P$, we leverage the node semantic representation 
$\textbf{W}$ to disentangle the interaction relation 
$\mathcal{R}$ into 
$K$ intent propagation matrices 
$\ddot{\mathcal{R}} \in \mathbb{R}^{K \times (|\mathcal{U}| + |\mathcal{I}|) \times (|\mathcal{U}| + |\mathcal{I}|)}$, which are normalized to obtain the normalized intent propagation matrix 
$\Bar{\mathcal{R}}$. 
The node representations 
$\textbf{M}$ are propagated through matrix 
$\Bar{\mathcal{R}}$, resulting in node representations 
$\textbf{N}$ after intent message propagation.
We implement this procedure according to the following steps.
\begin{itemize}[leftmargin=10pt]
\item \textbf{Step 1}:
Interaction Intensity Simulation.
In the context of implicit feedback environments, the interaction relations do not accurately reflect the intensity of interactions between nodes.
The degree-based adjacency matrix normalization method is relatively coarse.
Similar to DCCF \cite{renDisentangledContrastiveCollaborative2023}, to simulate the intensity of each interaction, we introduce a learnable intensity matrix $\tilde{\mathcal{R}}  \in \mathbb{R}^{(|\mathcal{U}| + |\mathcal{I}|) \times (|\mathcal{U}| + |\mathcal{I}|)}$ to encode the implicit relations between nodes.
The intensity matrix $\tilde{\mathcal{R}}$ is obtained through an element-wise multiplication of a mask matrix 
$\mathcal{T} \in \mathbb{R}^{(|\mathcal{U}| + |\mathcal{I}|) \times (|\mathcal{U}| + |\mathcal{I}|)}$ with the original interaction matrix 
$\mathcal{R}$, denoted as 
$\tilde{\mathcal{R}} = \mathcal{T} \odot \mathcal{R}$,
where $\odot$ represents an element-wise multiplication.
Each entry $\mathcal{T}_{u, v} \in [0, 1]$ reflects the interaction intensity between node $u$ and node $v$, with the value close to 1 indicating a stronger connection.
We derive $\mathcal{T}_{u,v}$ from the deep semantic representation of nodes $\textbf{W}$. 
Specifically, we use the cosine similarity between node embeddings \cite{chenIterativeDeepGraph2020} to measure the importance of interactions, which is defined as
\begin{equation}
\small
s(\textbf{w}_{u}, \textbf{w}_{v})
= \frac{
{\textbf{w}_{u}}^T \textbf{w}_{v}
}{
\Vert \textbf{w}_{u} \Vert_{2}
\Vert \textbf{w}_{v} \Vert_{2}
},
\end{equation}
where $\Vert \cdot \Vert_{2}$ represents the L2-norm method.
To ensure that the simulated intensity values are close to the range of [0,1], we utilize the following formula: 
$\mathcal{T}_{u,v}=(s(\textbf{w}_{u}, \textbf{w}_{v}) + 1) / 2$.

\item \textbf{Step 2}:
Construct Intent Message Propagation Matrix.
With these learnable intent embeddings $\tilde{\mathcal{R}}$, 
we implement the intent propagation matrix as
\begin{equation}
\small
\ddot{\mathcal{R}}^{k}_{u,v} = \tilde{\mathcal{R}}_{u,v}  \cdot
\frac{
{
( \textbf{w}_{u} \odot \textbf{w}_{v} )}^T \textbf{c}^k
}{
\sum_{o=1}^K ( \textbf{w}_{u} \odot \textbf{w}_{v} )^T \textbf{c}^o
},
\end{equation}
where $\ddot{\mathcal{R}}^{k}_{u,v}$ represents the weight of intent 
$k$ in the propagation process between node 
$u$ and node $v$, and $\odot$ represents an element-wise multiplication.
Then, by normalizing the weights of interactions between nodes under all intents, we construct the normalized intent propagation matrix as
\begin{equation}
\small
\Bar{\mathcal{R}}_{u,v}
= \frac{1}{K}
\sum_{k=1}^K
\frac{
\ddot{\mathcal{R}}_{u,v}
}{
\sum_{j \in \mathcal{N}_{\mathcal{R}}(u)} \ddot{\mathcal{R}}_{u,j}
}, 
\end{equation}
where 
$\Bar{\mathcal{R}}$ 
represents the normalized intent message propagation matrix, and $\mathcal{N}_{\mathcal{R}}(u)$ represents the set of first-order neighbors of node $u$ in graph $\mathcal{R}$.

\item \textbf{Step 3}:
Intent Message Propagation.
We construct the intent message propagation process based on the matrix $\Bar{\mathcal{R}}$, and 
% the specific process is as follows:
% \begin{equation}
% \small
% \textbf{N} = \Bar{\mathcal{R}} \cdot \textbf{M}.
% \end{equation}
the specific process is defined as
$\textbf{N} = \Bar{\mathcal{R}} \cdot \textbf{M}$.
Finally, we return the node representation $\textbf{N}$ generated by intent message propagation.

\end{itemize}

% \textcolor{blue}{After providing a clear definition of the Intent Message Propagation process, we disentangle the intent based on the direct interaction relation
% $\hat{\mathcal{A}}$ and the high-order interaction relation 
% $\hat{\mathcal{H}}$ between nodes, thereby defining the Intent Direct Message Propagation process and the Intent High-Order Message Propagation process.}

After providing the definition of the Intent Message Propagation, we perform intent disentangling and message propagation on the direct (high-order) interaction matrix 
$\hat{\mathcal{A}}$ ($\hat{\mathcal{H}}$), referred to as the Intent Direct (High-Order) Message Propagation.

\noindent\textbf{Intent Direct Message Propagation.}
In IPCCF, we integrate the node representations generated by the double helix message propagation framework. 
This enables us to effectively merge both the direct and high-order interaction relations between nodes. 
It facilitates a comprehensive understanding of the interaction intents between nodes.
This process can be defined as
\begin{equation}
\small
\textbf{W} = FCNN(Concat(\textbf{X}^{d},\textbf{X}^{h},\textbf{Y}^{d},\textbf{Y}^{h})),
\end{equation}
where $\textbf{W} \in \mathbb{R}^{(|\mathcal{U}|+|\mathcal{I}|) \times d}$ represents node representations that can reflect deep semantic information, 
FCNN represents a fully connected neural network, and Concat denotes the vector concatenation operation.
Subsequently, we invoke the intent direct message propagation procedure $P$, 
as
$\textbf{Z}^d = P(\textbf{W}, \hat{\mathcal{A}}, \textbf{F})$,
where
$\textbf{Z}^{d} \in \mathbb{R}^{(|\mathcal{U}| + |\mathcal{I}|) \times d}$ represents the node representations generated by intent direct message propagation.
Then, we update the 1st helical sequence by
\begin{equation}
\small
\textbf{F} = \textbf{X}^{d} + \textbf{Y}^{h} + \textbf{Z}^{d}.
\end{equation}

\noindent\textbf{Intent High-Order Message Propagation.}
We also invoke procedure $P$ to execute the high-order intent message propagation process, 
as
$\textbf{Z}^h = P(\textbf{W}, \hat{\mathcal{H}}, \textbf{S})$,
where
$\textbf{Z}^{h} \in \mathbb{R}^{(|\mathcal{U}|+|\mathcal{I}|) \times d}$ represents the node representations generated by intent high-order message propagation.
Then, we update the 2nd helical sequence by
\begin{equation}
\small
\textbf{S} = \textbf{X}^{h} + \textbf{Y}^{d} + \textbf{Z}^{h}.
\end{equation}

\subsubsection{Multilayer Message Propagation}

To capture deep collaborative filtering signals, we conduct message propagation based on GNNs across distinct layers.
The generation process of the embedding representation of the $l$-th layer is as 
\begin{equation}
\small
\textbf{E}_{l} = \textbf{E}_{l-1} + \textbf{F}_{l-1} + \textbf{S}_{l-1}.
\end{equation}
To reduce over-smoothing effects, we apply residual connections during the aggregation phase \cite{xiaHypergraphContrastiveCollaborative2022,renDisentangledContrastiveCollaborative2023}.

\subsection{Model Optimization}

\subsubsection{Recommendation Module}

After propagation through $L$ layers, we employ mean pooling to merge the embeddings from all layers, resulting in the final representation.
\begin{equation}
\small
\textbf{E}' = 
\frac{1}{L+1}
\sum_{l=0}^L \textbf{E}_l
\end{equation}
With the final representation, we predict the extent of interaction between user $u$ and item $i$ through the inner product,
% .
% \begin{equation}
% \small
% \hat{\mathcal{R}}_{u,i} = {\textbf{e}'_u}^T \textbf{e}'_i,
% \end{equation}
as
$\hat{\mathcal{R}}_{u,i} = {\textbf{e}'_u}^T \textbf{e}'_i$,
where $\hat{\mathcal{R}}_{u,i}$ represents the prediction score between user $u$ and item $i$.
To accurately capture user interaction information, we utilize the Bayesian Personalized Ranking (BPR) loss \cite{rendle2012bpr}, which is an objective function specifically designed for optimizing ranking in recommendation systems. 
Specifically, the BPR loss gives priority to observing user interactions, ensuring that the predicted scores for these interactions are higher than those for unobserved interactions. 
The BPR loss function is defined as 
\begin{equation}
\small
\mathcal{L}_{BPR} = 
- \frac{1}{|\mathcal{O}|}
\sum_{(u,i,j)}
\log 
\sigma
(\hat{\mathcal{R}}_{u,i} - \hat{\mathcal{R}}_{u,j}),
\end{equation}
where $\mathcal{O}$ denotes the training data, $i$ denotes an item that has interacted with user $u$, $j$ denotes an item that has not interacted with $u$, and $\sigma$ is the sigmoid function.

\subsubsection{Contrastive Learning}
Yu et al. \cite{yuAreGraphAugmentations2022} pointed out that directly constructing contrastive pairs within the same view, without the need for any augmentation operations, can yield better recommendation performance than using simple augmentation techniques.
This demonstrates the importance of other nodes within the same view in contrastive learning-based recommendation methods \cite{zhuAdaMCLAdaptiveFusion2023}. 
Following the method of supervised contrastive signals in \cite{wuSelfsupervisedGraphLearning2021}, we generate positive pairs using the embeddings of the same node in two views, while embeddings of different nodes are considered negative pairs. 
% By applying the InfoNCE loss \cite{chen2020simple}, we create contrastive self-supervision signals as
We use InfoNCE loss \cite{chen2020simple} to align node representations by maximizing the similarity of the same node and minimizing the similarity of different nodes across views.
The contrastive learning loss is formalized as
\begin{equation}
\small
\mathcal{I}(\textbf{T}', \textbf{T}'') =
\frac{1}{I}
\sum_{i=0}^I
\sum_{l=0}^L
-\log
\frac{
\exp(s(\textbf{t}'_{i,l}, \textbf{t}''_{i,l}) / \tau)
}{
\sum_{i'=0}^I \exp(s(\textbf{t}'_{i,l}, \textbf{t}''_{i',l}) / \tau)
},
\end{equation}
where $\textbf{t}'_{i,l}$ and $\textbf{t}''_{i,l}$ represent the embeddings of node $i$ in the $l$-th layer of the message propagation process in the first and second views, respectively, $s(\cdot, \cdot)$ represents the similarity function between nodes, and
$\tau$ is the temperature parameter.

\noindent\textbf{Double Helix Sequence Contrastive Learning.}
According to the previous study \cite{zhuAdaMCLAdaptiveFusion2023}, aligning with the fused view, which contains richer information, can enhance the effectiveness of contrastive learning.
Therefore, we align the representations $\textbf{F}$ and $\textbf{S}$ generated by our double helix sequence message propagation framework with the fused representation $\textbf{E}$ obtained through Eq.(16).
The contrastive learning loss is formalized as
\begin{equation}
\small
\mathcal{L}_{s} = \mathcal{I}(\textbf{E},\textbf{F})
+ \mathcal{I}(\textbf{E},\textbf{S}).
\end{equation}

\noindent\textbf{Propagation Process Contrastive Learning.}
We align the representations generated by the same type of message propagation process. This has two advantages: 
(1) Ensuring that the same type of message propagation generates the same node representation, thus ensuring that the node representation covers global structural information \cite{maoUltraGCNUltraSimplification2023}; 
(2) Using the structural relations between nodes to guide the disentangling process, providing direct supervision signals for the disentangling process.
The contrastive learning loss is formulated as
\begin{equation}
\small
\mathcal{L}_{p} = 
\mathcal{I}(\textbf{X}^d,\textbf{Y}^d)
+ \mathcal{I}(\textbf{X}^d,\textbf{Z}^d)
+ \mathcal{I}(\textbf{X}^h,\textbf{Y}^h)
+ \mathcal{I}(\textbf{X}^h,\textbf{Z}^h).
\end{equation}

\subsubsection{Independent Intent Constraint}
To enhance the disentangling process, we define the loss $\mathcal{L}_{i}$ 
as minimizing the similarity between $K$ intents. The process is defined as
\begin{equation}
\small
\mathcal{L}_{i} = \frac{2}{K \times (K-1)}
\sum_{i=1}^K \sum_{j=i+1}^K
s(\textbf{c}^i, \textbf{c}^j),
\end{equation}
where $s(\cdot)$ represents the similarity function between intents. 

\subsubsection{Joint Optimization}

The overall loss of IPCCF can be defined as
\begin{equation}
\small
\mathcal{L} = 
\mathcal{L}_{BPR} 
+ \lambda_1 \cdot \mathcal{L}_{s} 
+ \lambda_2 \cdot \mathcal{L}_{p}
+ \lambda_3 \cdot \mathcal{L}_{i}
+ \lambda_4 \cdot \Vert \Theta_1 \Vert^2_F 
+ \lambda_5 \cdot \Vert \Theta_2 \Vert^2_F,
\end{equation}
where $\lambda_1$, $\lambda_2$, $\lambda_3$, $\lambda_4$, and $\lambda_5$ are tunable weight parameters,
$\Theta_1 = \{ \textbf{E}_0 \}$,
$\Theta_2 = \{ \textbf{C} \}$ are tunable vector embeddings
and $\Vert \cdot \Vert^2_F$ represents the Frobenius norm method.

\subsection{Model Analysis}

\subsubsection{Analysis of the design rationale of IPCCF}

Methods such as LightGCN \cite{heLightGCNSimplifyingPowering2020} and NCL \cite{linImprovingGraphCollaborative2022}  have demonstrated the effectiveness of structure-based message propagation. However, these methods are coarse-grained, overlooking the intensity and intent of node interactions, which leads to node representations that fail to reflect users' actual preferences. While intent-disentanglement-based approaches offer a more detailed message propagation process, existing methods \cite{wangDisentangledGraphCollaborative2020,renDisentangledContrastiveCollaborative2023,zhang2024exploring} focus primarily on disentangling direct interactions between nodes, neglecting a more comprehensive graph-structural perspective. Additionally, the disentanglement process lacks explicit supervision signals, resulting in biases and model overfitting.

To address the above issues, IPCCF has been designed with the following aspects:
\begin{itemize}[leftmargin=10pt]

\item 
We extract high-order relations between nodes, which bring the following benefits: (1) Message propagation based on high-order relations can improve node representations and provide strong semantic support for the model's disentangling operations. (2) High-order relations create efficient message propagation paths that work alongside direct interactions, expanding the range of considerations for disentangling operations and greatly improving their accuracy.

\item 
We propose a double-helix message propagation framework that integrates various interaction types. This framework effectively mines semantic information, enriching the disentanglement process and enhancing the model's understanding of interaction intents.

% We extract high-order interaction relations between nodes and propose a double helix message propagation framework that effectively integrates different types of interaction information between nodes. 
% This framework can deeply mine the semantic information of nodes in the graph structure, providing richer semantic support for the disentanglement process and enhancing the model’s understanding of interaction intents between nodes.

\item 
We propose an intent propagation mechanism that simulates interaction intensity and disentangles node intents based on deep semantic information. By aggregating node intents within the graph, the model can evaluate the disentanglement process from a more comprehensive structural perspective, thereby improving accuracy.

% We propose an intent propagation mechanism. In this mechanism, based on the deep semantic information of nodes, we simulate the intensity of interactions between nodes and disentangle the interaction intents between them. 
% Through intent propagation, we aggregate the intents of nodes on the graph structure. 
% This enables the model to assess the rationality of the disentanglement process from a more comprehensive graph structure perspective, thereby enhancing the accuracy of disentanglement.

\item 
Through contrastive learning, we align node representations from structure-based and intent-based message propagation. This approach offers three advantages: (1) providing supervision signals for disentanglement, reducing bias, and enhancing resistance to overfitting; (2) considering both interaction intensity and intents, which resolves ambiguities in structure-based propagation; (3) aligning node representations at different stages, enabling the model to capture global structural information.

% Through contrastive learning, we align node representations generated from the structure-based and intent-based message propagation processes. This approach has three advantages: i) It provides direct supervision signals for disentanglement, thereby reducing disentanglement bias and enhancing the model’s ability to resist overfitting; ii) During disentanglement, we consider both interaction intensity and intents between nodes. Through contrastive learning, we effectively address ambiguity in structure-based message propagation regarding these aspects. These outcomes guide initial node representations to achieve more reasonable distributions in latent space; iii) Based on previous research \cite{maoUltraGCNUltraSimplification2023}, we align node representations generated at different stages of the message propagation process, enabling the model to capture global structural information.

\item 
We simplify the intent message propagation process by omitting the specific message propagation details under each intent. This ensures that the model achieves effective disentanglement while maintaining high efficiency.

\end{itemize}

% In IPCCF, we combine the strengths of existing GCNs and disentanglement methods \cite{linImprovingGraphCollaborative2022,renDisentangledContrastiveCollaborative2023}. 
% Our model has the following advantages over existing methods:

\subsubsection{Comparison with Existing Methods}
\begin{itemize}[leftmargin=10pt]

\item \textbf{Comparison with existing GCNs methods.}
In the process of building message propagation, we emphasize the intensity and intent of interactions between nodes instead of focusing solely on structural information \cite{heLightGCNSimplifyingPowering2020}.
% In the process of building message propagation, we go beyond just structural information \cite{heLightGCNSimplifyingPowering2020}. 
% We focus on the intensity and intent of interactions between nodes.
This approach aims to minimize the gap between node representation and their actual distribution.

\item \textbf{Comparison with graph attention networks.}
Compared with methods like GAT \cite{velickovicGraphAttentionNetworks2018} and HAN \cite{wangHeterogeneousGraphAttention2021}, our approach goes beyond using simple node pair similarities. We focus on deep semantic relations between nodes and adjust message propagation weights from a broader perspective.

\item \textbf{Comparison with existing disentangled methods.}
% We effectively address the accuracy issues in existing disentanglement models \cite{wangDisentangledGraphCollaborative2020,renDisentangledContrastiveCollaborative2023,zhang2024exploring} caused by the limited scope of consideration.
% Additionally, we mitigate bias and model overfitting caused by the lack of direct supervision during the disentanglement process.
Compared with methods such as DCCF \cite{renDisentangledContrastiveCollaborative2023} and BIGCF \cite{zhang2024exploring}, we effectively address the accuracy issues caused by the limited scope of consideration, while also mitigating bias and model overfitting due to the lack of direct supervision during the disentanglement process.
Compared with KGIN \cite{wang2021learning}, both our model and KGIN achieve multi-intent disentanglement in node interactions. 
However, KGIN aggregates all messages without classifying them by intent type, truncating the propagation of same-intent messages and hindering the effective evaluation of the disentanglement's rationality within the comprehensive graph structure.
Compared with DGCF \cite{wangDisentangledGraphCollaborative2020}, both our model and DGCF implement message propagation under a single intent. However, DGCF lacks guidance from the graph structure during the disentanglement process, which causes the disentanglement objective to focus on maximizing the final recommendation performance on the training data, rather than accurately capturing the true interaction intent between nodes. This can lead to model bias and overfitting.

\end{itemize}

\subsubsection{Time Complexity Analysis}

We analyze the time complexity of various components in IPCCF from the following aspects:
i) The shallow message propagation process takes $\mathcal{O}(L \times (|\mathcal{A}| + |\mathcal{H}|) \times d)$ time, where $L$ is the number of layers in the graph neural network message propagation,  $|\mathcal{A}|$ is the number of edges in the graph, $|\mathcal{H}|$ is the number of edges in the high-order propagation relations, and $d$ is the dimension of nodes.
ii) The deep message propagation has the same time complexity as the shallow message propagation.
iii) The intent message propagation process takes  $\mathcal{O}( (L+K) \times (|\mathcal{A}| + |\mathcal{H}|) \times d)$  time, where $K$ is the number of intents.
iv) The cost of contrastive learning is $\mathcal{O} (L \times B \times (|\mathcal{U}| + |\mathcal{I}|) \times d) $, where $B$ is the number of users/items included in a single mini-batch.

\section{EVALUATION}
To demonstrate the effectiveness of IPCCF, we have conducted experiments on three popular datasets to answer the following questions:
% \begin{itemize}[leftmargin=25pt]
    % \setlength{\itemsep}{0pt}
    % \setlength{\parsep}{0pt}
    % \setlength{\parskip}{0pt}

\begin{itemize}[leftmargin=10pt]

\item \textbf{RQ1:} How does IPCCF perform compared to other recommendation algorithms in different datasets?

\item \textbf{RQ2:} Do the key components of IPCCF positively contribute to its performance?

\item \textbf{RQ3:} Does IPCCF maintain its superior performance on sparse datasets?

\item \textbf{RQ4:} What is the impact of the number of latent intents on performance?

\item \textbf{RQ5:} Can IPCCF mitigate the issue of over-smoothing in node representations?

\item \textbf{RQ6:} How does our IPCCF perform w.r.t training efficiency?

\end{itemize}

\subsection{Experimental Setting}

\begin{table}[t]
\setlength{\abovecaptionskip}{5pt}
\setlength{\belowcaptionskip}{-5pt}
  \centering
  \scriptsize
  \caption{Properties of Datasets}
  \label{table:simu-set}
  % \resizebox{\linewidth}{!}{
  \resizebox{8cm}{!}{
    \begin{tabular}{C{1.8cm}C{1.0cm}C{1.0cm}C{1.5cm}C{1.0cm}}
    \toprule
    Datasets  & $\#$Users &  $\#$Items & $\#$Interactions & Density  \\ \midrule
    Gowalla    & 50,821  &  57,440  & 1,172,425  & $4.0e^{-4}$       \\ 
    Amazon-book    & 78,578  & 77,801  & 2,240,156  & $3.7e^{-4}$      \\ 
    Tmall & 47,939  & 41,390  & 2,357,450  & $1.2e^{-3}$     \\  
    \bottomrule
    \end{tabular}
}
\end{table}

\subsubsection{Datasets}

To assess the performance of IPCCF, we have conducted experiments on three public datasets: Gowalla, Amazon-book, and Tmall. 
% \textcolor{blue}{
% These datasets differ in scale, field, and sparsity, and are the same as the datasets used in DCCF \cite{renDisentangledContrastiveCollaborative2023} and BIGCF \cite{zhang2024exploring}.}
Table II provides the details of these datasets.
\begin{itemize}[leftmargin=10pt]
\item \textbf{Gowalla.} 
This dataset collects check-in interactions between users and various locations, collected from the Gowalla platform, focusing on mobility tracking \cite{renDisentangledContrastiveCollaborative2023,zhang2024exploring}.

\item \textbf{Amazon-book.} 
% This dataset contains user evaluations of
% book-related products on Amazon.
This dataset, provided by Amazon, contains large-scale book reviews, including user evaluations and interaction records related to book products on the Amazon platform \cite{renDisentangledContrastiveCollaborative2023,wangDisentangledGraphCollaborative2020,zhang2024exploring,caiLightGCLSimpleEffective2023}.

\item \textbf{Tmall.} 
% It comprises consumer purchasing activities from the e-commerce platform Tmall.
This dataset, provided by Alibaba, contains consumer purchasing activities from the e-commerce platform Tmall \cite{renDisentangledContrastiveCollaborative2023,zhang2024exploring,caiLightGCLSimpleEffective2023}.

\end{itemize}

\subsubsection{Evaluation Metrics}
To reduce the bias in sampling item instances, we followed a full-rank protocol for all items \cite{heLightGCNSimplifyingPowering2020,wangDisenHANDisentangledHeterogeneous2020}, which measures the accuracy of our recommendations.
% We utilized two commonly used ranking-based metrics, Recall@N and NDCG@N, to evaluate the performance of all approaches.
To comprehensively evaluate the performance of our model, we assess all methods across three dimensions: relevance, coverage, and ranking quality. These evaluation dimensions have also been used in previous studies \cite{wangDisenHANDisentangledHeterogeneous2020,heLightGCNSimplifyingPowering2020,zhang2024exploring}. For this purpose, we employ three widely used ranking-based evaluation metrics: Precision@K, Recall@K, and NDCG@K.

\begin{itemize}[leftmargin=10pt]
\item \textbf{Precision@K}. 
This metric measures the proportion of relevant items in the top-K recommendations relative to all recommended items, reflecting recommendation relevance.
% This metric measures the proportion of relevant items in the top-K recommendations, reflecting their relevance.
% This metric measures the proportion of relevant items in the top-K recommendations relative to all recommended items, reflecting the relevance of the model's recommendations.
\item \textbf{Recall@K}. 
This metric measures the proportion of relevant items in the top-K recommendations relative to all relevant items, reflecting recommendation coverage.
% This metric measures the proportion of relevant items in the top-K recommendations relative to all relevant items, reflecting the coverage of the model's recommendations.
\item \textbf{NDCG@K}. This metric evaluates the ranking quality of the top-K recommendations by considering both the relevance and position of recommendation items, reflecting the model’s ability to rank items based on their relevance.
\end{itemize}

\subsubsection{Baselines}

Besides IPCCF, we adopt several state-of-the-art methods as baselines. 
These baselines are categorized into three groups: recommendation with GNNs, self-supervised learning for recommendation, and disentangled multi-intent recommender systems.

\noindent\textbf{Recommendation with GNNs.}
\begin{itemize}[leftmargin=10pt]
\item \textbf{NGCF} \cite{wangNeuralGraphCollaborative2019}.
This method integrates user-item interaction information into embeddings. 
It achieves the modeling of bipartite graph high-order connectivity.

\item \textbf{LightGCN} \cite{heLightGCNSimplifyingPowering2020}.
This method maintains effectiveness while simplifying the structure and reducing model complexity, leading to easier training and improved performance.
% This method maintains effectiveness while simplifying the structure and reducing the complexity of the model. This leads to easier model training and better training performance.

\end{itemize}

\noindent\textbf{Self-supervised learning for recommendation.}
\begin{itemize}[leftmargin=10pt]
\item \textbf{SLRec} \cite{yaoSelfsupervisedLearningLargescale2021}. 
This method introduces a self-supervised learning framework for large-scale recommendation, addressing label sparsity by learning hidden feature connections.
% This method introduces a self-supervised learning (SSL) framework for large-scale item recommendation.
% The framework mainly solves the issue of sparse labels by better learning the hidden connections between item features.

\item \textbf{SGL-ED/ND}
\cite{wuSelfsupervisedGraphLearning2021}.
This method enhances representation learning by adding a self-supervised contrastive task using data augmentation, specifically edge or node drop.
% This method improves the learning of user/item representations using GNNs. It achieves this by adding a self-supervised contrastive learning task. 
% This task uses data augmentation, specifically edge drop (ED) or node drop (ND).

\item \textbf{HCCF} 
\cite{xiaHypergraphContrastiveCollaborative2022}.
This method uses a hypergraph-enhanced cross-view contrastive learning framework to jointly capture local and global collaborative relations.

\item \textbf{LightGCL}
\cite{caiLightGCLSimpleEffective2023}.
This method uses Singular Value Decomposition, extracting information from interactions and injecting global context into contrastive learning representations.
% This method uses Singular Value Decomposition (SVD) to enhance graphs, extracting useful information from user-item interactions and injecting global collaborative context into contrastive learning representations.

\end{itemize}

\noindent\textbf{Disentangled multi-intent recommender systems.}
\begin{itemize}[leftmargin=10pt]
\item \textbf{DisenGCN}
\cite{maDisentangledGraphConvolutional}.
This method proposes a neighbor routing mechanism to dynamically select nodes for information extraction and convolution.

\item \textbf{DisenHAN}
\cite{wangDisenHANDisentangledHeterogeneous2020}.
This method proposes a disentangled heterogeneous graph attention network to learn user/item representations from heterogeneous information aspects.

\item \textbf{CDR} \cite{chenCurriculumDisentangledRecommendation}.
This method uses collaborative filtering, dynamic routing, and curriculum learning to denoise and explore intent relations.

\item \textbf{DGCF}
\cite{wangDisentangledGraphCollaborative2020}.
This method models intent distribution, improves the graph, and encourages intent independence, yielding separate representations.

\item \textbf{DGCL}
\cite{liDisentangledContrastiveLearning}.
This method uses a factor-wise discrimination target to obtain distinct node representations and predict user-item interactions with inner products.

\item \textbf{DCCF} \cite{renDisentangledContrastiveCollaborative2023}.
This method uses global context learning for disentangled representations, extracting latent factors and reducing noise.
% This method uses global context learning to achieve disentangled representations, extracting finer latent factors and reducing augmentation-induced noise. It introduces a cross-view contrastive learning task with an adaptive interaction mask generator for enhancement.

\item \textbf{BIGCF} \cite{zhang2024exploring}.
This method examines user-item interactions causally, introducing individual intent for personal preferences and collective intent for overall awareness.
% This method carefully examines user-item interactions from a causal perspective and introduces the concepts of individual intent representing personal preferences and collective intent representing overall awareness.

\end{itemize}

\subsubsection{Hyperparameter Setting}
We implemented IPCCF using PyTorch and utilized the Adam optimizer during training, with the learning rate set to 0.001. The number of message propagation layers $L$ was set to 2.
The parameters $\lambda_1$ and $\lambda_2$ were fine-tuned within the range of \{$10^{-2}$, $2 \times 10^{-2}$, $4 \times 10^{-2}$, $6 \times 10^{-2}$, $8 \times 10^{-2}$, $10^{-1}$, $2 \times 10^{-1}$, $4 \times 10^{-1}$\}, and the parameters $\lambda_3$, $\lambda_4$, and $\lambda_5$ were fine-tuned within the range of \{$2.5 \times 10^{-6}$, $5 \times 10^{-6}$, $10^{-5}$, $2.5 \times 10^{-5}$, $5 \times 10^{-5}$, $10^{-4}$, $2.5 \times 10^{-4}$, $5 \times 10^{-4}$, $10^{-3}$, $2.5 \times 10^{-3}$, $5 \times 10^{-3}$, $10^{-2}$\}.
During the extraction of high-order interactions between nodes, for all datasets, the filter threshold $Q$ was set to 5, and the filter threshold $\eta$ was set to 0.8.
The number $K$ of latent intents was selected from the set $\{4, 8, 16, 32\}$. 
% Optimal performance is achieved with 8 prototypes on dataset Gowalla, and 16 prototypes on datasets Amazon-book and Tmall.
To evaluate baseline performance under fair conditions, the latent embedding dimension $d$ and batch size for all comparative methods were set to 32 and 10,240, respectively.
Based on our experiments, the model achieves the best performance with
$\lambda_2$ set to $10^{-1}$, $\lambda_3$ set to $5 \times 10^{-3}$, $\lambda_4$ set to $2.5 \times 10^{-5}$,
$\lambda_5$ set to $10^{-5}$, $K$ set to 8, 
$\lambda_1$ set to $8 \times 10^{-2}$ for the Gowalla dataset, and $\lambda_1$ set to $10^{-1}$ for the Amazon-book and Tmall datasets.
To enable replication of our model, detailed hyperparameter configurations are provided in the source code at https://github.com/rookitkitlee/IPCCF.

\begin{table*}
\setlength{\abovecaptionskip}{5pt}
\setlength{\belowcaptionskip}{-5pt}
\caption{Recommendation performance of all compared methods on different datasets in terms of Precision@20/40, Recall@20/40 and NDCG@20/40. The bold and the underlined values indicate the best and the second best values, respectively.}

\scriptsize
\centering
% \resizebox{\linewidth}{!}{
\resizebox{17cm}{!}{
\begin{tabular}{C{1.5cm}C{0.6cm}C{0.6cm}C{0.6cm}C{0.6cm}C{0.6cm}C{0.6cm}C{0.6cm}C{0.6cm}C{0.6cm}C{0.6cm}C{0.6cm}C{0.6cm}C{0.6cm}C{0.6cm}C{0.6cm}C{0.6cm}C{0.6cm}C{0.6cm}}
\toprule
\multirow{2}*{Algorithm}  & \multicolumn{6}{c}{Gowalla}    & \multicolumn{6}{c}{Amazon-book}   & \multicolumn{6}{c}{Tmall}     \\
\cmidrule(lr){2-7}   \cmidrule(lr){8-13}  \cmidrule(lr){14-19}  

& P@20 & P@40 & R@20  & R@40 &  N@20  & N@40  
& P@20 & P@40 & R@20  & R@40 &  N@20  & N@40 
& P@20 & P@40 & R@20  & R@40 &  N@20  & N@40  \\
\midrule

NGCF   
& 0.0189 & 0.0141 
& 0.1413 & 0.2072 & 0.0813 & 0.0987   
& 0.0184 & 0.0153 
& 0.0532 & 0.0866 & 0.0388   & 0.0501 
& 0.0117 & 0.0104 
& 0.0420 & 0.0751 & 0.0250  & 0.0365  \\

LightGCN   
& 0.0238 & 0.0175 
& 0.1799 & 0.2577 & 0.1053 & 0.1255  
& 0.0252 & 0.0202 
& 0.0732 & 0.1148 & 0.0544  & 0.0681 
& 0.0152 & 0.0123 
& 0.0555 & 0.0895 & 0.0381  & 0.0499  \\

\midrule

SLRec   
& 0.0204 & 0.0148 
& 0.1529 & 0.2200 & 0.0926 & 0.1102    
& 0.0188 & 0.0156 
& 0.0544 & 0.0879 & 0.0374 & 0.0490  
& 0.0149 & 0.0122 
& 0.0549 & 0.0888 & 0.0375 & 0.0492   \\

SGL-ED   
& 0.0240 & 0.0173 
& 0.1809 & 0.2559 & 0.1067 & 0.1262  
& 0.0268 & 0.0212 
& 0.0774 & 0.1204 & 0.0578 & 0.0719   
& 0.0157 & 0.0126 
& 0.0574 & 0.0919 & 0.0393 & 0.0513   \\

SGL-ND   
& 0.0241 & 0.0176 
& 0.1814 & 0.2589 & 0.1065 & 0.1267    
& 0.0248 & 0.0198 
& 0.0722 & 0.1121 & 0.0542 & 0.0674  
& 0.0152 & 0.0122
& 0.0553 & 0.0885 & 0.0379 & 0.0494   \\

HCCF    
& 0.0240 & 0.0178 
& 0.1818 & 0.2601 & 0.1061 & 0.1265 
& 0.0283 & 0.0226 
& 0.0824 & 0.1282 & 0.0625 & 0.0776  
& 0.0171 & 0.0135 
& 0.0623 & 0.0986 & 0.0425 & 0.0552   \\

LightGCL    
& 0.0242 & 0.0177 
& 0.1825 & 0.2601 & 0.1077 & 0.1280 
& 0.0289 & 0.0228 
& 0.0836 & 0.1280 & 0.0643 & 0.0790   
& 0.0173 & 0.0134 
& 0.0632 & 0.0971 & 0.0444 & 0.0562   \\

\midrule

DisenGCN 
& 0.0185 & 0.0137 & 0.1379 & 0.2003 & 0.0798 & 0.0961   
& 0.0167 & 0.0138 & 0.0481 & 0.0776 & 0.0353 & 0.0451 
& 0.0117 & 0.0095 & 0.0422 & 0.0688 & 0.0285  & 0.0377  \\

DisenHAN   
& 0.0192 & 0.0141 
& 0.1437 & 0.2079 & 0.0829 & 0.0997    
& 0.0186 & 0.0153 
& 0.0542 & 0.0865 & 0.0407 & 0.0513 
& 0.0116 & 0.0096
& 0.0416 & 0.0682 & 0.0283  & 0.0376  \\

CDR
& 0.0184 & 0.0132
& 0.1364 & 0.1943 & 0.0812 & 0.0963    
& 0.0196 & 0.0157
& 0.0564 & 0.0887 & 0.0419   & 0.0526 
& 0.0143 & 0.0116
& 0.0520 & 0.0833 & 0.0356  & 0.0465  \\

DGCF 
& 0.0236 & 0.0170
& 0.1784 & 0.2515 & 0.1069 & 0.1259   
& 0.0237 & 0.0190
& 0.0688 & 0.1073 & 0.0513 & 0.0640   
& 0.0149 & 0.0119
& 0.0544 & 0.0867 & 0.0372 & 0.0484   \\

DGCL 
& 0.0236 & 0.0168 & 0.1793 & 0.2483 & 0.1067 & 0.1247    
& 0.0233 & 0.0186 & 0.0677 & 0.1057 & 0.0506 & 0.0631   
& 0.0145 & 0.0116 & 0.0526 & 0.0845 & 0.0359 & 0.0469   \\

% DCCF     
% & \underline{0.1876} & \underline{0.2644} & \underline{0.1123} & \underline{0.1323}    
% & \underline{0.0889} & \underline{0.1343} & \underline{0.0680} & \underline{0.0829}   
% & \underline{0.0668} & \underline{0.1042} & \underline{0.0469} & \underline{0.0598}   \\

DCCF     
& 0.0252 & 0.0181 & 0.1876 & 0.2644 & 0.1123 & 0.1323   
& 0.0306 & 0.0237 & 0.0889 & 0.1343 & 0.0680 & 0.0829 
& 0.0183 & 0.0142 & 0.0668 & 0.1042 & 0.0469 & 0.0598   \\

BIGCF     
& \underline{0.0280} & \underline{0.0199} 
& \underline{0.2086} & \underline{0.2883} & \underline{0.1242} & \underline{0.1450}    
& \underline{0.0338} & \underline{0.0258} 
& \underline{0.0989} & \underline{0.1468} & \underline{0.0761} & \underline{0.0918} 
& \underline{0.0205} & \underline{0.0158} 
& \underline{0.0755} & \underline{0.1167} & \underline{0.0535} & \underline{0.0680}   \\

\midrule
\textbf{IPCCF(Ours)}       
& \textbf{0.0297} & \textbf{0.0209}
& \textbf{0.2178} & \textbf{0.3002} & \textbf{0.1310} & \textbf{0.1527} 
& \textbf{0.0376} & \textbf{0.0277}
& \textbf{0.1106} & \textbf{0.1562} & \textbf{0.0894} & \textbf{0.1043} 
& \textbf{0.0216} & \textbf{0.0163}
& \textbf{0.0803} & \textbf{0.1210} & \textbf{0.0579} & \textbf{0.0721} \\
\midrule

\textit{Improvement} 
& +6.07\% & +5.03\%
& +4.41\% & +4.13\% & +5.48\% & +5.31\% 
& +11.24\% & +7.36\%
& +11.83\% & +6.40\% & +17.48\% & +13.62\%
& +5.37\% & +3.16\%
& +6.36\% & +3.68\% & +8.22\% & +6.03\%
\\
\bottomrule
\end{tabular}
}
\end{table*}

\subsection{Experimental Results and Analysis}

\subsubsection{Performance Comparison (RQ1)}
In this section, we report the performance comparison between IPCCF and all baselines across three datasets. 
The results are shown in Table III, with the best results highlighted in bold and the second-best results underlined for emphasis.
From the experimental results, we can observe that: 
(1) Compared to GNN-based methods \cite{heLightGCNSimplifyingPowering2020,wangNeuralGraphCollaborative2019}, most SSL-based methods \cite{xiaHypergraphContrastiveCollaborative2022,caiLightGCLSimpleEffective2023} and some disentangled recommendation methods \cite{liDisentangledContrastiveLearning,renDisentangledContrastiveCollaborative2023} can achieve better recommendation performance. 
(2) IPCCF achieves the best performance, significantly outperforming all baselines in two metrics across three datasets. 
Across different Top $K$ ranges $(K=20,40)$ in terms of Precision, Recall and NDCG, on average, our model surpasses the best baseline by $5.55\%$, $4.27\%$ and $5.40\%$ on Gowalla, $9.30\%$, $9.12\%$ and $15.55\%$ on Amazon-Book, and $4.27\%$, $5.02\%$ and $7.13\%$ on Tmall, respectively.
This represents a significant increase in the recommendation, especially considering that the best baseline, BIGCF \cite{zhang2024exploring}, has already made significant improvements over the second-best baseline. 
This strongly demonstrates the effectiveness of our model.

The reasons for these experimental results are as follows:
(1) GNN-based methods \cite{heLightGCNSimplifyingPowering2020,wangNeuralGraphCollaborative2019} fail to address the data sparsity of interaction and the entangled interaction intent problems in real-world recommendation scenarios. 
By constructing interaction graphs from multiple perspectives, SSL-based methods \cite{xiaHypergraphContrastiveCollaborative2022,wuSelfsupervisedGraphLearning2021} can effectively alleviate the data sparsity problem. 
For example, LightGCL \cite{caiLightGCLSimpleEffective2023} addresses the data sparsity problem effectively by injecting global collaborative relations, providing additional guidance for message propagation between nodes.
Disentanglement methods \cite{liDisentangledContrastiveLearning,renDisentangledContrastiveCollaborative2023}  explore the actual intents behind node interactions, capturing user preferences more accurately. 
For example, the current state-of-the-art disentangling methods, DCCF \cite{renDisentangledContrastiveCollaborative2023} and BIGCF \cite{zhang2024exploring}, explore the intent representations of nodes. This allows the generation of node representations to go beyond the structural information of the graph, thereby better uncovering user preferences.
Thus, SSL-based methods and disentanglement methods significantly outperform GNN-based methods in recommendation scenarios.
(2)
Compared to other disentanglement models \cite{renDisentangledContrastiveCollaborative2023,zhang2024exploring}, IPCCF introduces a double helix message propagation framework to extract the deep semantic information of nodes.
This enhances the model's understanding of interaction intents between nodes.
It also introduces an intent message propagation mechanism to incorporate graph structures into the consideration scope of disentanglement. 
Through contrastive learning, IPCCF aligns node representations from structure-based and intent-based message propagation. 
It provides direct supervision signals to enhance disentanglement and reduce biases in the process.
Consequently, IPCCF achieves better recommendation performance.

\begin{table}
\setlength{\abovecaptionskip}{5pt}
\setlength{\belowcaptionskip}{-5pt}
  \centering
  \scriptsize
  \caption{Ablation study on key components of IPCCF (measured by Recall@20 and NDCG@20) on different datasets.}
  % \resizebox{\linewidth}{!}{
  \resizebox{8cm}{!}{
    \begin{tabular}{C{0.85cm}C{0.75cm}C{0.75cm}C{0.75cm}C{0.75cm}C{0.75cm}C{0.75cm}}
    \toprule
    \multirow{2}*{Algorithm} 
    & \multicolumn{2}{c}{Gowalla} 
    & \multicolumn{2}{c}{Amazon-book}
    & \multicolumn{2}{c}{Tmall} \\ 
    \cmidrule{2-3} \cmidrule{4-5} \cmidrule{6-7}
    & R@20   & N@20  & R@20   & N@20 & R@20   & N@20        \\  \midrule

    \textbf{w/o ho}   & 0.1926 & 0.1156   & 0.0921 & 0.0708    & 0.0681   & 0.0479     \\
    \textbf{w/o dp}   & 0.2086 & 0.1264   & 0.1050 & 0.0845    & 0.0760   & 0.0546     \\
    \textbf{w/o he}   & 0.2078 & 0.1251   & 0.1043 & 0.0844    & 0.0757   & 0.0543     \\
    \textbf{w/o ip}   & 0.1747 & 0.1012   & 0.0858 & 0.0650    & 0.0674   & 0.0476     \\

    \textbf{w/o spc}   & 0.1978 & 0.1184   & 0.0998 & 0.0787    & 0.0701   & 0.0492     \\
    
    \textbf{w/o sc}   & 0.2148 & 0.1290   & 0.1096 & 0.0887    & 0.0798   & 0.0574     \\
    \textbf{w/o pc}   & 0.2121 & 0.1271   & 0.1030 & 0.0817    & 0.0726   & 0.0516     \\

    \textbf{w/o pcd}   & 0.2136 & 0.1280   & 0.1062 & 0.0854    & 0.0755   & 0.0541     \\

    \textbf{w/o pci}   & 0.2154 & 0.1291   & 0.1086 & 0.0879    & 0.0788   & 0.0565     \\
    
     \midrule
    \textbf{IPCCF}    & 0.2178 & 0.1310   & 0.1106 &  0.0894   & 0.0803   & 0.0579   \\ \bottomrule 

    \end{tabular}
}
\end{table}

\subsubsection{Ablation Study (RQ2)}
In this section, we validate the effectiveness of each component within the IPCCF model through nine variants:

\begin{itemize}[leftmargin=10pt]

\item \textbf{w/o ho}: 
Eliminate the message propagation process in IPCCF that is constructed based on high-order interactions between nodes.

\item \textbf{w/o dp}: 
Eliminate the deep message propagation component.

\item \textbf{w/o he}: 
Eliminate cross-transmission of different message propagation processes and switch to sequential transmission of the same type.

\item \textbf{w/o ip}: 
Eliminate the intent message propagation component.

\item \textbf{w/o spc}:
Eliminate all contrastive learning losses from the model, including those for double-helix sequences and propagation processes.

\item \textbf{w/o sc}:
Eliminate the double-helix sequence contrastive learning loss.

\item \textbf{w/o pc}:
Eliminate the contrastive learning loss of the propagation process.

\item \textbf{w/o pcd}:
Eliminate aligning node representations for shallow and deep message propagations in Propagation Process Contrastive Learning.

\item \textbf{w/o pci}:
Eliminate aligning node representations for shallow and intent message propagations in Propagation Process Contrastive Learning.

\end{itemize}

The results of the ablation study are presented in Table IV, where we focus solely on two metrics: Recall@20 and NDCG@20.
Similar results were observed across other evaluation metrics as well.
From the experimental results, it is evident that:
\begin{itemize}[leftmargin=10pt]

\item 
The experimental results of variants \textbf{w/o ho}, \textbf{w/o dp}, and \textbf{w/o he} confirm that IPCCF effectively integrates both direct and indirect interactions between nodes, extracting deep semantic representations of nodes within the graph structure. These measures significantly enhance recommendation performance.

\item 
The experimental results of variant \textbf{w/o ip} demonstrate that disentanglement significantly impacts recommendation performance in IPCCF. This is because our disentangling process simulates the intent and intensity of interactions between nodes, refines the interactions between nodes, and enhances node representation.

\item 
The experimental results of variants \textbf{w/o sc} and \textbf{w/o pc} demonstrate that in IPCCF, aligning node representations generated from different message propagation processes enables us to effectively mitigate the differences between the distributions of various views and the actual data.
This helps to address the sparsity and noise problems of data in recommendation scenarios.
Furthermore, experimental results also demonstrate that providing direct supervision signals through contrastive learning for the disentangling process can reduce its bias and improve recommendation performance.

\item 
We categorize the contrastive learning modules in IPCCF into three types, corresponding to variants \textbf{w/o sc}, \textbf{w/o pcd}, and \textbf{w/o pci}, and analyze them along with variant \textbf{w/o spc}. 
We find that eliminating any single contrastive learning module has a relatively small impact on the model's performance. 
This is because the modules overlap in functionality, so eliminating one does not fully capture its role. 
Further analysis of variants \textbf{w/o sc}, \textbf{w/o pcd}, and \textbf{w/o pci} reveals that the contrastive learning module for shallow and deep message propagation has a stronger impact on performance, as it controls the model's initial stages and therefore exerts a greater influence on overall performance.

\end{itemize}

\begin{figure}[t]
\setlength{\abovecaptionskip}{0pt}
\setlength{\belowcaptionskip}{-5pt}
\centering
% \includesvg[scale=0.38]{Figure/P1.svg}
\includegraphics[scale=0.38]{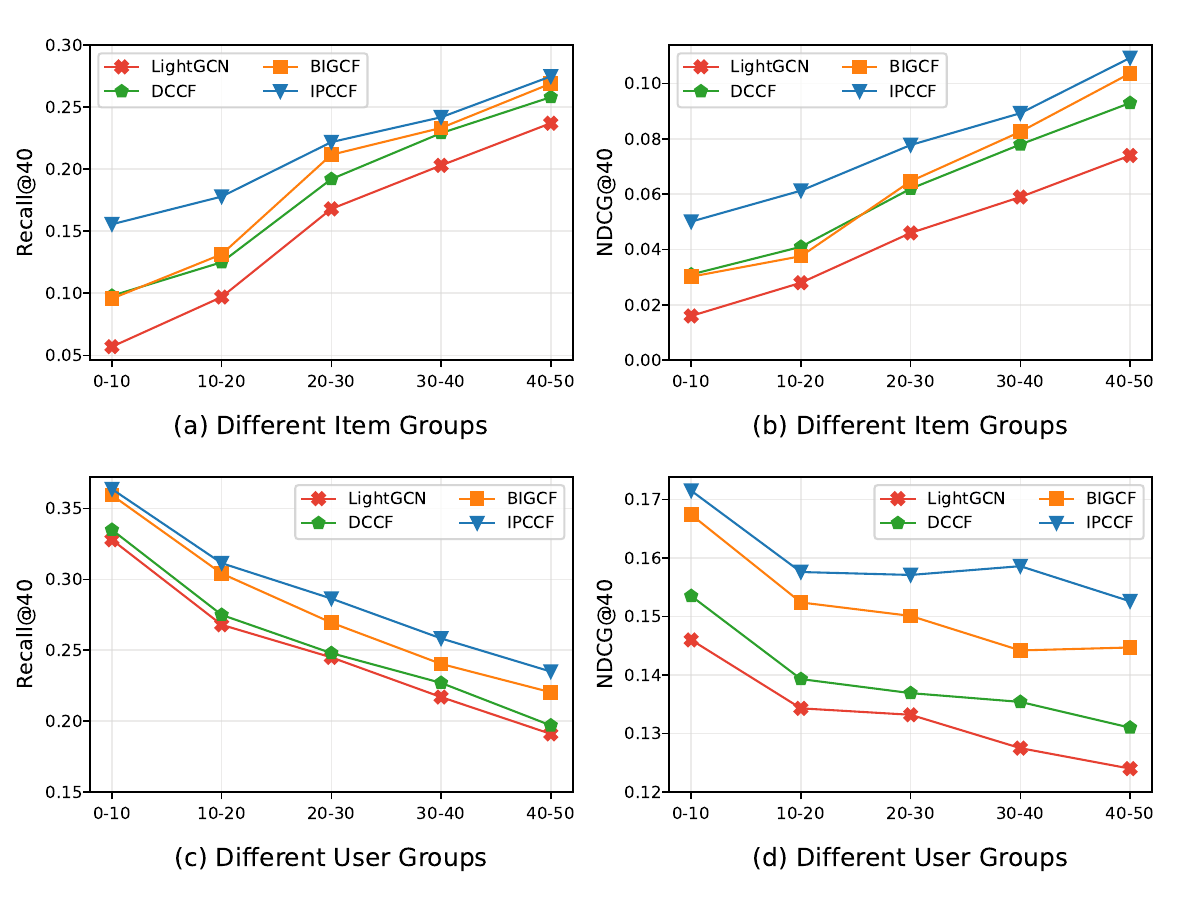}
\caption{Performance comparison w.r.t. data sparsity over
different user/item groups on the Gowalla dataset.}
\label{fig:label}
\end{figure}

\begin{figure}[t]
\setlength{\abovecaptionskip}{0pt}
\setlength{\belowcaptionskip}{-5pt}
\centering
% \includesvg[scale=0.38]{Figure/P3.svg}
\includegraphics[scale=0.38]{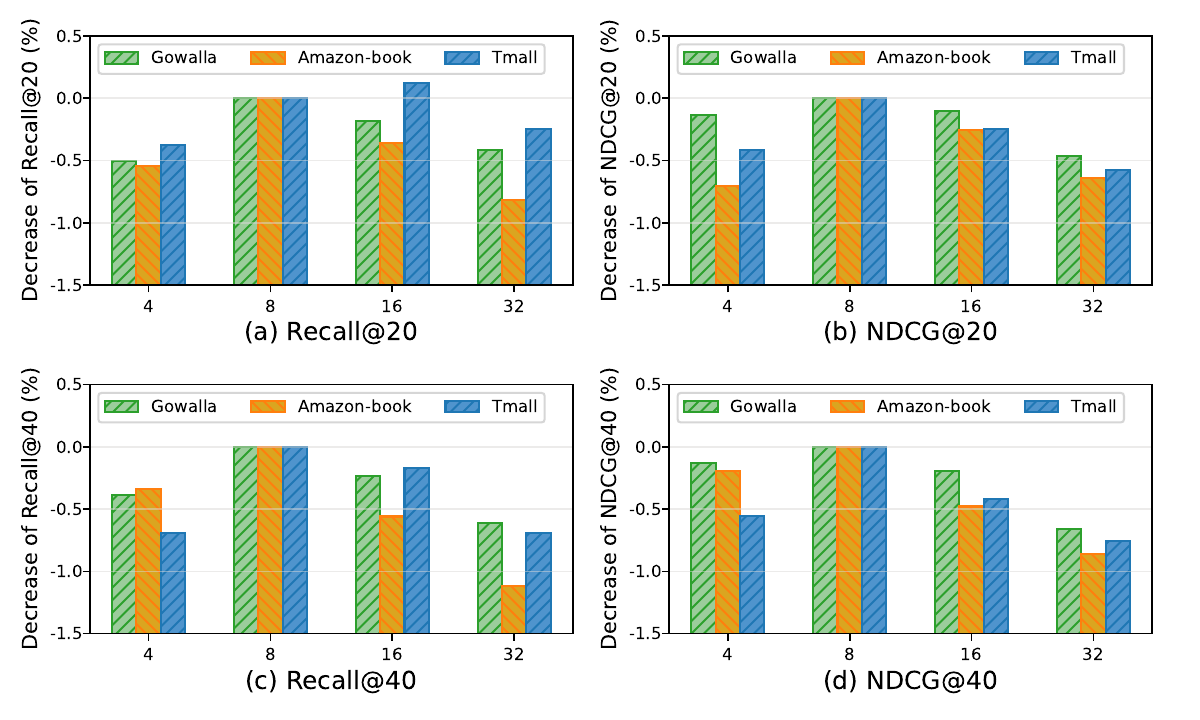}
\caption{Performance w.r.t the number of latent intents.}
\label{fig:label}
\end{figure}

\begin{table}
\setlength{\abovecaptionskip}{5pt}
\setlength{\belowcaptionskip}{-5pt}
  \centering
  \scriptsize
  \caption{The embedding smoothness on the Amazon-book and Tmall datasets measured by MAD metric (the smaller the MAD indicates more obvious over-smoothing phenomenon).}
  % \resizebox{\linewidth}{!}{
  \resizebox{8cm}{!}{
    \begin{tabular}{C{0.9cm}|C{0.9cm}C{0.9cm}C{0.9cm}C{0.9cm}C{0.9cm}C{0.9cm}}
    \toprule
     Type
    & LightGCN
    & DisenGCN
    & DGCL
    & DCCF
    & BIGCF
    & IPCCF
    \\ 
    \midrule
   
    & \multicolumn{6}{c}{Amazon-book}   \\  \midrule
    User     & 0.984 & 0.961   & 0.980 & 0.999 & 0.999 & \textbf{0.999}     \\ 
    Item     & 0.944 & 0.986   & 0.989 & 0.990 & 0.994   & \textbf{0.996}   \\ \midrule

    & \multicolumn{6}{c}{Tmall}   \\  \midrule
    User     & 0.910 & 0.876   & 0.897 & \textbf{0.999} & 0.999 & 0.986     \\ 
    Item     & 0.927 & 0.992   & 0.920 & \textbf{0.998} & 0.995  & 0.995   \\

    \bottomrule 

    \end{tabular}

   }
\end{table}

\subsubsection{In-Depth Analysis of IPCCF (RQ3, RQ4 \& RQ5)}\   

\noindent\textbf{Performance w.r.t. Data Sparsity.}
To validate the robustness of IPCCF against the issue of data sparsity, we categorized users and items into different groups based on the number of interactions between them. We evaluated the recommendation performance for each group. The experimental results are presented in Fig. 3.

These results show that IPCCF's recommendation performance consistently surpasses all baseline models, regardless of the sparsity levels of users and items. 
According to the current research \cite{renDisentangledContrastiveCollaborative2023}, utilizing contrastive learning methods has proven to be effective in addressing the data sparsity problem within recommendation systems.
In IPCCF, we have created a multi-dimensional contrastive learning process based on helical sequences and the propagation process.
Compared to existing models \cite{renDisentangledContrastiveCollaborative2023,zhang2024exploring}, the comparative view constructed by IPCCF offers a finer granularity, providing more precise supervision signals during the learning process. 
Therefore, IPCCF solves the data sparsity problem more effectively than existing models.

\noindent\textbf{Impact of the Number of Intents.}
We trained the model with intent parameters chosen from the set $\{4,8,16,32\}$ to investigate the impact of the number of potential intents on model performance. 
The experimental results are shown in Fig. 4. From these results, we obtained two key observations: (1) The best recommendation performance on all datasets was achieved with 8 intents. 
(2) Changes in the number of intents had a minimal impact on the model's recommendation performance.

This result can be attributed to the fact that in the IPCCF model, intents are represented by learnable parameter embeddings.
This allows the model to adjust its learning strategy according to the number of intents. Such adaptability means that the model can find the optimal or near-optimal learning path across different numbers of intents.

\noindent\textbf{Robustness of IPCCF in Alleviating Over-Smoothing.}
We validated the effectiveness of IPCCF in mitigating over-smoothing by calculating the Mean Average Distance (MAD) \cite{chenMeasuringRelievingOversmoothing2019,renDisentangledContrastiveCollaborative2023} for all users and items. 
It measures the smoothness of the graph by calculating the average distance between node representations \cite{chenMeasuringRelievingOversmoothing2019}. 
A higher MAD value indicates a greater average distance between node representations, suggesting that the generated node representations are more resistant to smoothing.
We compared it with several representative baseline methods, such as LightGCN \cite{heLightGCNSimplifyingPowering2020}, DisenGCN \cite{maDisentangledGraphConvolutional}, DGCL \cite{liDisentangledContrastiveLearning}, DCCF \cite{renDisentangledContrastiveCollaborative2023},
and BIGCF \cite{zhang2024exploring}. 
Following the DCCF approach, we normalized all embeddings before computing MAD. The experimental results are shown in Table V. 
From these results, we can see that IPCCF remains highly competitive in alleviating over-smoothing compared to other methods.

This success is due to the use of a double helix structure during the disentangling process, which deeply explores the semantic information of nodes, and the application of an intent message propagation model to broaden the consideration scope of disentanglement.
By this means, even nodes with the same neighbors can maintain significant differences. 
This helps reduce the problem of over-smoothing in node representations.

\begin{table}
\setlength{\abovecaptionskip}{5pt}
\setlength{\belowcaptionskip}{-5pt}
  \centering
  \scriptsize
  \caption{Computational cost evaluation in terms of per-epoch training time (seconds).}
  % \resizebox{\linewidth}{!}{
  \resizebox{8cm}{!}{
    \begin{tabular}{C{1.8cm}C{1.0cm}C{1.0cm}C{1.0cm}C{1.0cm}}
    \toprule

    Model   & LightGCN  &  DCCF & BIGCF   & IPCCF    \\  \midrule

    Gowalla     & \textbf{10.5s} & 18.7s  & 12.0s &   16.5s  \\ 
    Amazon-book & \textbf{11.9s} & 24.6s  & 12.8s &   20.1s  \\
    Tmall       & \textbf{11.5s} & 24.5s  & 12.7s &   19.2s  \\
    
   \bottomrule 

    \end{tabular}

   }
\end{table}

% \begin{table}
%   \centering
%   \scriptsize
%   \caption{Computational cost evaluation in terms of per-epoch training time (seconds) on Gowalla, Amazon-book, and Tmall data.}
%   \resizebox{\linewidth}{!}{
%     \begin{tabular}{C{1.7cm}C{0.8cm}C{0.8cm}C{0.8cm}C{0.8cm}C{0.8cm}C{0.8cm}}
%     \toprule

%     Model    &DGCF  & DGCL & DCCF  & BIGCF & IPCCF    \\  \midrule

%     Gowalla     & 25.0s & 9.2s   & 13.4s  & 6.8s  &   13.2s  \\ 
%     Amazon-book & 49.8s & 12.2s   & 19.2s  & 7.5s &   17.6s  \\
%     Tmall      &  52.2s & 12.0s   & 19.0s  & 7.4s &   15.9s  \\
    
%    \bottomrule 

%     \end{tabular}

%    }
% \end{table}

% \begin{table}
%   \centering
%   \scriptsize
%   \caption{Computational cost evaluation in terms of per-epoch training time (seconds) on Gowalla, Amazon-book, and Tmall data.}
%   \resizebox{\linewidth}{!}{
%     \begin{tabular}{C{1.7cm}C{0.8cm}C{0.8cm}C{0.8cm}C{0.8cm}C{0.8cm}C{0.8cm}}
%     \toprule

%     Model    &DGCF  & DGCL & DCCF  & BIGCF & IPCCF    \\  \midrule

%     Gowalla     & 30.3s & 14.6s   & 18.7s  & 12.0s &   18.5s  \\ 
%     Amazon-book & 55.2s & 17.6s   & 24.6s  & 12.8s &   23.0s  \\
%     Tmall      &  57.7s & 17.5s   & 24.5s  & 12.7s &   21.5s  \\
    
%    \bottomrule 

%     \end{tabular}

%    }
% \end{table}

\begin{figure}[t]
\setlength{\abovecaptionskip}{0pt}
\setlength{\belowcaptionskip}{-5pt}
\centering
% \includesvg[scale=0.39]{Figure/P4.svg}
\includegraphics[scale=0.39]{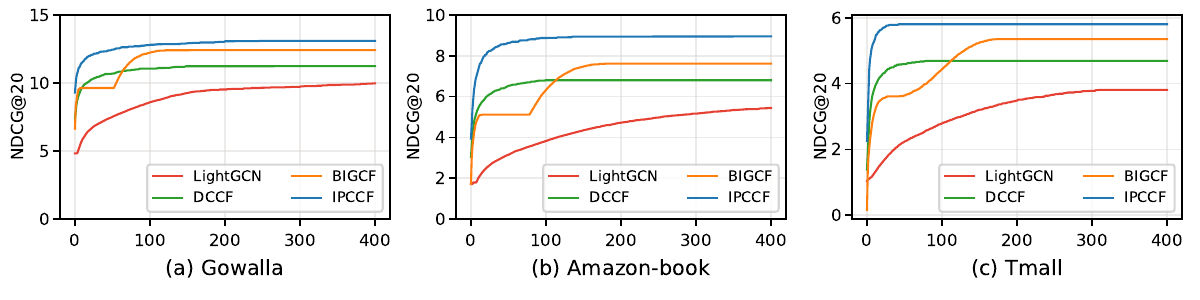}
\caption{Performance comparison over different training epochs.}
\label{fig:label}
\end{figure}

\begin{table}
\setlength{\abovecaptionskip}{5pt}
\setlength{\belowcaptionskip}{-5pt}
  \centering
  \scriptsize
  \caption{The proportion of time required for IPCCF to achieve the same experimental results as other models.}
  % \resizebox{\linewidth}{!}{
  \resizebox{8cm}{!}{
    \begin{tabular}{C{1.5cm}C{1.5cm}C{1.5cm}C{1.5cm}}
    \toprule

    Model   & LightGCN  &  DCCF & BIGCF       \\  \midrule

    Gowalla      & $3.16\%$  & $7.95\%$ &   $49.68\%$  \\ 
    Amazon-book  & $3.42\%$  & $7.35\%$ &   $14.67\%$  \\
    Tmall        & $2.54\%$  & $5.98\%$ &   $11.52\%$  \\
    
   \bottomrule 

    \end{tabular}

   }
\end{table}

\subsubsection{Model Training Efficiency Study (RQ6)}

In this section, we explore the model efficiency of IPCCF in terms of training computational cost across all datasets. 
The experiments were conducted on a server equipped with an Intel(R) Xeon(R) Platinum 8358P CPU and an NVIDIA RTX 3090 GPU.

\noindent\textbf{Cost Comparison for Each Training Epoch.}
We measured the time cost per training epoch, with each epoch consisting of 40,960 data divided into 40 batches. The results are shown in Table VI.

Compared to other models, we have enhanced the extraction of deep semantic information from nodes, improving the model’s ability to handle disentangling processes. However, experimental results show that the IPCCF model maintains comparable time consumption due to our simplified intent message propagation, which omits the propagation process under each intent.

\noindent\textbf{Performance Comparison Over Different Training Epochs.}
We compared the performance of different models over training epochs with a learning rate of 0.001 and NDCG@20 as the evaluation metric. The results are shown in Fig. 5.

Experimental results show that, compared to other models, IPCCF quickly improves performance in early training stages. This is due to (1) the double helix message propagation framework, integrating diverse node interactions, and (2) the intent message propagation processes, incorporating graph structures into disentanglement. These designs enrich node representation and provide richer supervision signals during training, reducing parameter oscillations.

\noindent\textbf{Tradeoff Between Performance and Cost.}
We evaluated the cost per training epoch and the performance across models over different epochs, calculating IPCCF's cost to achieve the peak performance of other models. The results are shown in Table VII.

Experimental results show that IPCCF achieves the same performance as LightGCN, DCCF, and BIGCF with just $3.04\%$, $7.09\%$, and $25.29\%$ of their training time, respectively. It offers the best performance-cost tradeoff, effectively handling model updates with new interaction records, highlighting its practical value.
For example, in e-commerce, user behavior changes during promotions. Users focus more on discounts and popular items instead of regular-priced products. Traditional models update slowly and may recommend outdated items. In contrast, IPCCF, with its optimal performance-cost tradeoff, enables more frequent model retraining. It quickly learns user preferences during promotions and adjusts recommendations to be more relevant.

\begin{figure}[t]
\setlength{\abovecaptionskip}{0pt}
\setlength{\belowcaptionskip}{-5pt}
\centering
% \includesvg[scale=0.44]{Figure/P5.svg}
\includegraphics[scale=0.34]{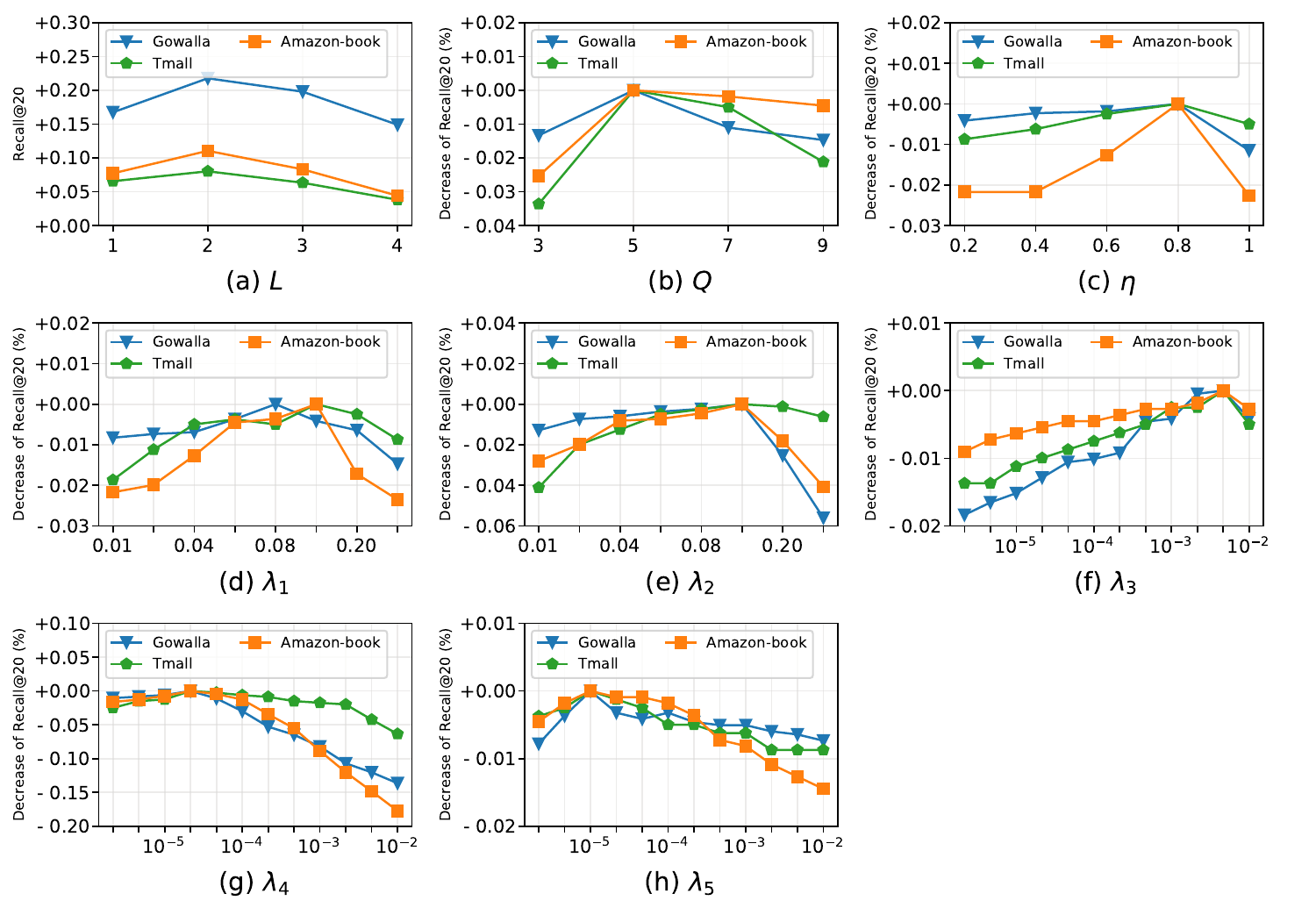}
\caption{Hyperparameter study of the IPCCF.}
\label{fig:label}
\end{figure}

\subsubsection{Hyperparameter Analysis}
In this section, we evaluate the impact of the number of message propagation layers $L$, filtering thresholds $Q$ and $\eta$, and tunable weight parameters $\lambda_1$, $\lambda_2$, $\lambda_3$, $\lambda_4$, and $\lambda_5$ on the model's performance.
The detailed range of hyperparameter values can be found in Section IV.A.4.
We present the experimental results based on the Recall@20 metric for datasets Gowalla, Amazon-book, and Tmall. 
Similar conclusions can be reached for other evaluation metrics. 
The experimental results for hyperparameter variations across different datasets are shown in Fig. 6.

\noindent\textbf{Impact of the Number of Message Propagation Layers.}
The experimental results on the impact of the number of message propagation layers are shown in Fig. 6(a).
These results demonstrate that as the number of message propagation layers increases, the model's performance first improves and then declines, with the best performance achieved when $L = 2$. 
This is because when the number of message propagation layers is too small, the node representation can only capture local structural information, limiting the model's performance. 
Conversely, when the number of layers is too large, the node representation becomes overly smooth, which also impairs the model's performance.

\noindent\textbf{Impact of the filter thresholds.}
The experimental results on the impact of the filter thresholds are shown in Fig. 6(b-c).
As can be seen from Fig. 6(b), the model's performance initially improves and then declines as more high-order relations are extracted between homogeneous nodes. 
This is because extracting such relations enhances the model's understanding of the graph structure, thereby improving performance. However, excessive extraction can introduce noise, ultimately impairing the model's performance.
This reasoning also explains the experimental results shown in Fig. 6(c). 
Filter threshold $\eta$ complements threshold $Q$, enabling the algorithm to extract more reliable high-order relations. 
As $\eta$ decreases, more relations can be extracted. However, if $\eta$ is set too small, it may introduce noise, thereby impairing the model's performance.

\noindent\textbf{Impact of the tunable weight parameters.}
The experimental results on the impact of the tunable weight parameters are shown in Fig. 6(d-h).
These
results demonstrate that when the model achieves optimal performance, all weight parameters are less than 1. 
This is because different weights correspond to different optimization objectives, which differ from the main objective of the recommendation system. 
To minimize interference, the weight values are kept small. Additionally, the weight values are positively correlated with the similarity between their corresponding objectives and the recommendation system's main objective: the higher the similarity, the larger the weight. 
Therefore, at optimal performance, the values of $\lambda_1$ and $\lambda_2$ are greater than those of $\lambda_3$, $\lambda_4$, and $\lambda_5$.

\begin{figure}[t]
\setlength{\abovecaptionskip}{0pt}
\setlength{\belowcaptionskip}{-5pt}
\centering
% \includesvg[scale=0.34]{Figure/F4.svg}
\includegraphics[scale=0.34]{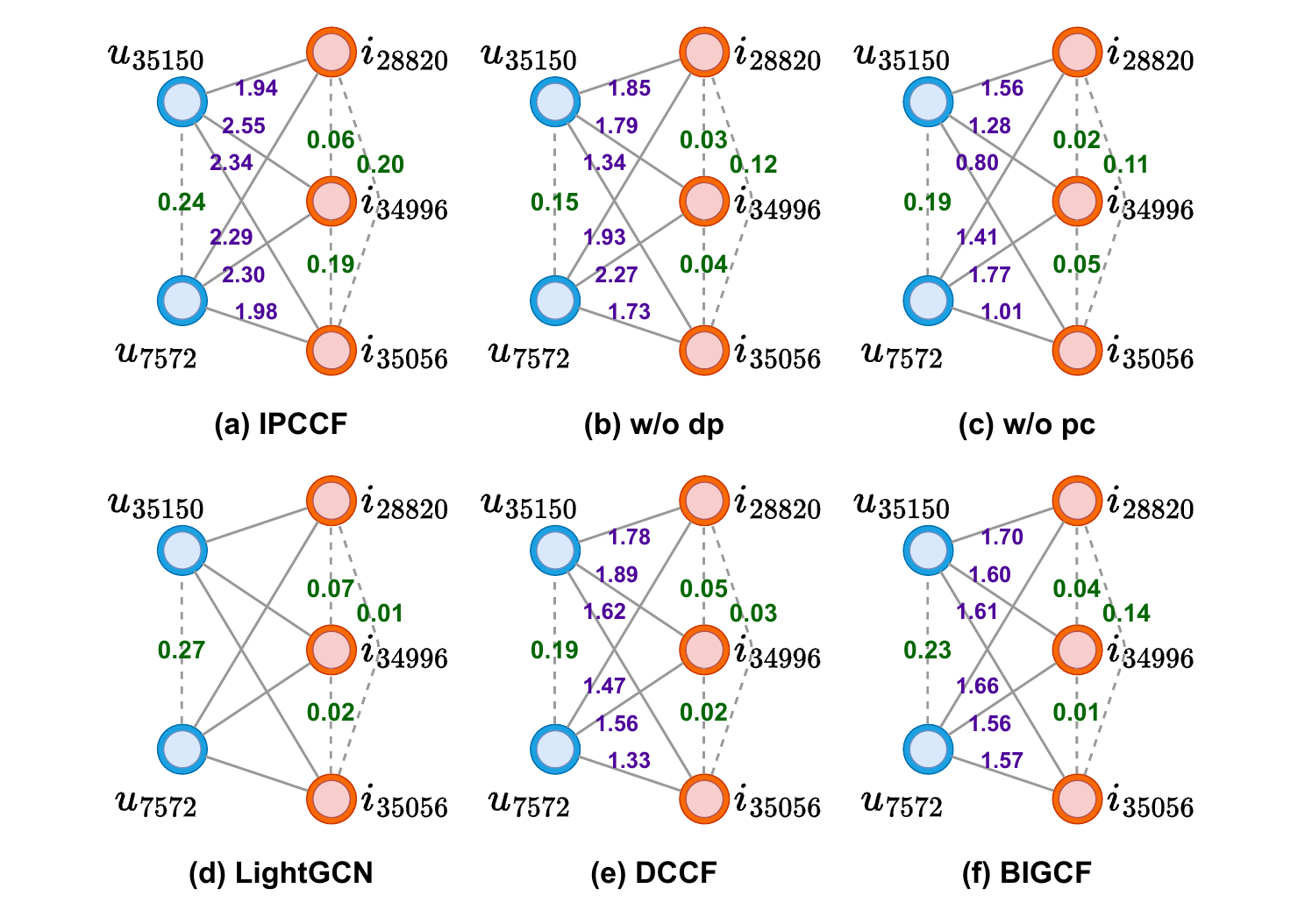}
\caption{A case study.}
\label{fig:label}
\end{figure}

\subsubsection{Case Study}
We extracted a subgraph from the Tmall dataset consisting of users $u_{35056}$ and $u_{35150}$, items $i_{28820}$, $i_{34996}$, and $i_{7572}$, and their interactions.
This subgraph was used as an example to analyze the disentangling results of the IPCCF model and its variants \textbf{w/o dp} and \textbf{w/o pc}, as well as LightGCN, DCCF, and BIGCF.
We analyzed the similarity between homogeneous nodes using the inner product and evaluated the distribution of interaction intents between heterogeneous nodes using information entropy. 
Since LightGCN does not include a disentangling component, we omit the analysis of its interaction intents. 
As DCCF and BIGCF perform disentangling at the node level, to ensure fairness, we first compute the intent distributions for the two interacting nodes and subsequently calculate the square root of the product of the two nodes' distributions.
The experimental results are shown in Fig. 7.

% The experimental results demonstrate that the IPCCF algorithm performs better than the variants \textbf{w/o dp} and \textbf{w/o pc} regarding similarity among homogeneous nodes and diversity of interaction intents among heterogeneous nodes.
The experimental results demonstrate that the IPCCF algorithm performs better than the variant \textbf{w/o dp}, the variant \textbf{w/o pc}, DCCF, and BIGCF regarding the similarity among homogeneous nodes and the diversity of interaction intents among heterogeneous nodes.
This is because IPCCF extracts deep semantic information, enhancing its understanding of interaction relations between nodes. 
Incorporating graph structures into the scope of disentanglement expands the considerations of disentanglement.
Direct supervision signals are provided for the disentanglement process through contrastive learning.
Therefore, IPCCF can explore the diversity of intent interactions among nodes from a more comprehensive graph structural perspective. 
This enhances the similarity of representations for homogeneous nodes and effectively avoids model overfitting caused by the pursuit of recommendation optimization goals.
Moreover, we found that LightGCN shows higher similarity among some homogeneous nodes than other models. 
This is because it lacks intent-level supervision in generating node representations, which leads to over-smoothing.

\noindent\textbf{Summary:}
The above experiments show that IPCCF is an effective recommendation algorithm. 
We tested the performance of IPCCF on three different datasets. 
On average, compared to other algorithms, our model improved the Precision metric by $6.37\%$, the Recall metric by $6.14\%$ and the NDCG metric by $9.36\%$.
Through ablation studies, we confirmed the effectiveness of each component of IPCCF. 
We demonstrated IPCCF's robustness across data sparsity, intent number, and mitigating over-smoothing. 
The training efficiency of our model was studied to ensure that good training efficiency is achieved while maintaining excellent performance.
Additionally, we analyzed IPCCF's disentanglement results through a case study.
% In addition, we analyzed the disentanglement results of IPCCF through a case study.

\section{CONCLUSION AND FUTURE WORK}

In this paper, we have proposed the \textbf{I}ntent \textbf{P}ropagation \textbf{C}ontrastive \textbf{C}ollaborative \textbf{F}iltering (IPCCF) algorithm, where we designed a double helix message propagation framework to better integrate the structural relations of nodes.
Meanwhile, we have developed an intent message propagation method that incorporates graph structures into the disentangling process, expanding the consideration scope of disentanglement.
Additionally, through contrastive learning techniques, we aligned the representations generated based on structures and those generated based on intents.
This can reduce biases and enhancing the model’s resistance to overfitting.
Extensive experimental results demonstrate the superiority of IPCCF.
% demonstrate that IPCCF significantly outperforms existing collaborative filtering models.
In the future, we plan to extract richer semantic information of nodes from text and knowledge graphs to further enhance the disentanglement effectiveness.

\section{ACKNOWLEDGMENTS}
This work is sponsored by 
the National Natural Science Foundation of China (Nos. 62172249, 62202253, 62172249, 62472441), 
the Natural Science Foundation of Shandong Province (Nos. ZR2021MF092, ZR2021QF074, ZR2022MF326)
and
the China Scholarship Council (Grant No. 202308370301).

\bibliographystyle{IEEEtran}
\bibliography{ipccfref}

\vspace{-12mm}
\begin{IEEEbiography}[{\includegraphics[width=1in,height=1.25in,clip,keepaspectratio]{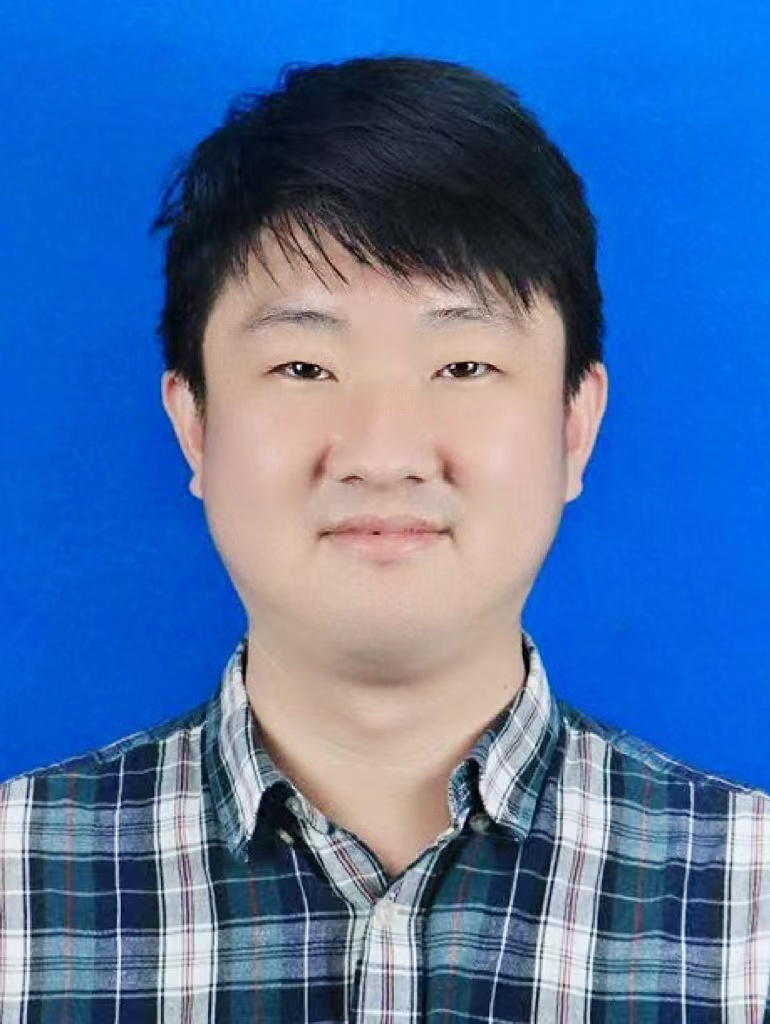}}]{Haojie Li}
is currently working toward
the Ph.D. degree in Qingdao University of Science and Technology (QUST), Qingdao, China. His current research interests include graph mining, graph representation learning, and recommender systems.
\end{IEEEbiography}

\vspace{-12mm}
\begin{IEEEbiography}[{\includegraphics[width=1in,height=1.25in,clip,keepaspectratio]{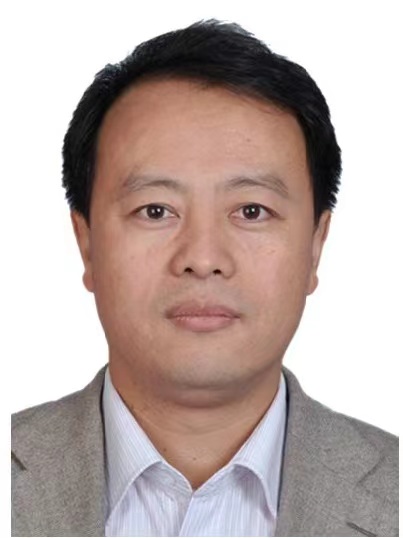}}] 
{Junwei Du} 
is a professor at Qingdao University of Science and Technology (QUST), Qingdao, China. He is the excutive dean of the School of Data Science of QUST and a distinguish member of CCF. He received his Ph.D. degree in computer theory and science from Tongji University, Shanghai, China. His current research interest includes graph representation learning, recommendation algorithm, and natural language processing. He has published more than 100 papers.
\end{IEEEbiography}

\vspace{-12mm}
\begin{IEEEbiography}[{\includegraphics[width=1in,height=1.25in,clip,keepaspectratio]{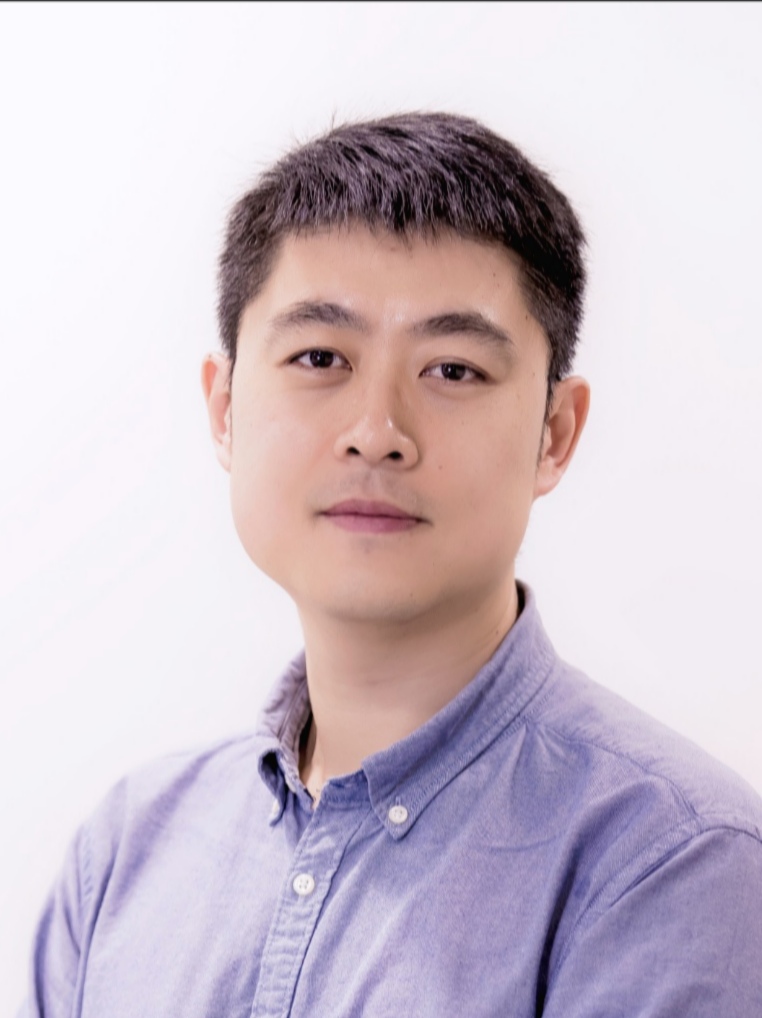}}]{Guanfeng Liu}
(Senior Member, IEEE) received the
PhD degree in computer science from Macquarie
University in 2013. He is currently with the School
of Computing, Macquarie University, Sydney, Australia. His research interests include graph database,
trust computing, and Graph Neural Netowrks. He
has published more than 200 papers on international
journals and conferences including TKDE, TSC,
TKDD, TWEB, AAAI, IJCAI, WWW and ICDE.
\end{IEEEbiography}

\vspace{-12mm}
\begin{IEEEbiography}[{\includegraphics[width=1in,height=1.25in,clip,keepaspectratio]{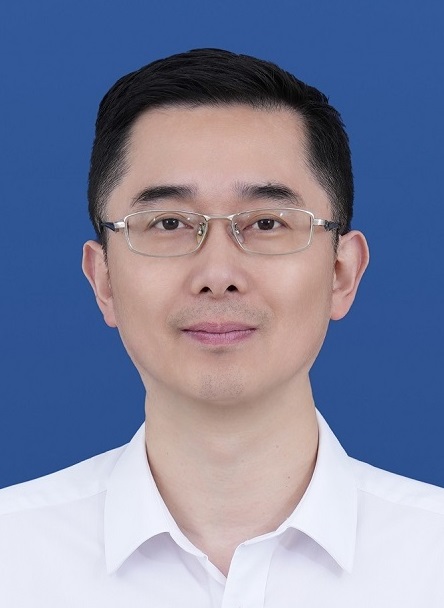}}]{Feng Jiang}
received the Ph.D. degree in computer software from the Institute of Computing Technology, Chinese Academy of Sciences, China, in 2007. He is currently a professor of computer science in Qingdao University of Science and Technology, Qingdao, China. He is a member of the Granular Computing and Knowledge Discovery Professional Committee of the Chinese Artificial Intelligence Society. His research interests include machine learning, rough set theory, and data mining.
\end{IEEEbiography}

% \vspace{-12mm}
% \begin{IEEEbiography}[{\includegraphics[width=1in,height=1.25in,clip]{Photo/WY_photo.jpg}}]{Yan Wang} (Senior Member, IEEE) received his PhD degree in computer science and technology from Harbin Institute of Technology (HIT), China in 1996. He is currently a Professor in the School of Computing, Macquarie University, Australia. He has published a number of research papers in international conferences including AAAI, AAMAS, ICDE, IJCAI, NeurIPS, SIGIR, WWW, and journals including CSUR, TIST, TKDE, TPDS, TSC and TWEB. His research interests cover trust management, recommender systems and service computing. 
% \end{IEEEbiography}

\vspace{-12mm}
\begin{IEEEbiography}[{\includegraphics[width=1in,height=1.25in,clip]{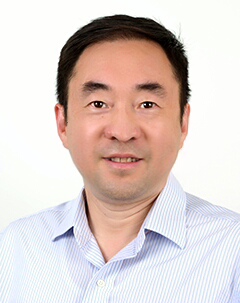}}]{Yan Wang} (Senior Member, IEEE) received his
PhD degree in computer science and technology
from Harbin Institute of Technology (HIT), China
in 1996. He is currently a Professor in the School
of Computing, Macquarie University, Australia. He
has published a number of research papers in international conferences including AAAI, AAMAS,
ICDE, IJCAI, NeurIPS, SIGIR, WWW, and journals
including CSUR, TIST, TKDE, TPDS, TSC and
TWEB. His research interests cover trust management, recommender systems and service computing.
\end{IEEEbiography}

\vspace{-12mm}
\begin{IEEEbiography}[{\includegraphics[width=1in,height=1.25in,clip]{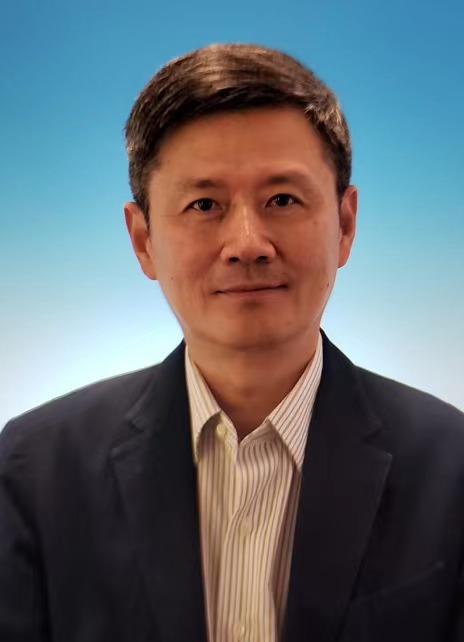}}]{Xiaofang Zhou} (Fellow, IEEE) received the BSc
and MSc degrees in computer science from Nanjing
University, China, and the PhD degree in computer
science from the University of Queensland, Australia,
in 1984, 1987, and 1994, respectively. He is a chair
professor of computer science with The Hong Kong
University of Science and Technology. His research
interests include spatial and multimedia databases,
high performance query processing, web information
systems, data mining, bioinformatics, and e-research.
\end{IEEEbiography}

\end{document}